\documentclass[lineno,authoryear]{FLO_v1}%

\usepackage{graphicx}
\usepackage{upgreek}
\usepackage{multicol,multirow}
\usepackage{amsmath,amssymb,amsfonts}
\usepackage{mathrsfs}
\usepackage{amsthm}
\usepackage[figuresright]{rotating}
\usepackage{appendix}
\usepackage[authoryear]{natbib}
\usepackage{ifpdf}
\usepackage[T1]{fontenc}
\usepackage{newtxtext}
\usepackage{newtxmath}
\usepackage{textcomp}
\usepackage{xcolor}
\usepackage{comment}

\usepackage[colorlinks,allcolors=blue]{hyperref}
\definecolor{jourcolor}{cmyk}{1,0.57,0.01,0.38}
\hypersetup{
    colorlinks,%
    citecolor=jourcolor,%
    filecolor=jourcolor,%
    linkcolor=jourcolor,%
    urlcolor=jourcolor
}

\theoremstyle{definition}

\articletype{RESEARCH ARTICLE}

\begin{document}

%\title[Wall-displacement based resolvent analysis]{Wall-displacement based resolvent analysis for prediction of flow response to wall roughness}

\title[Wall-displacement based resolvent analysis]{Receptivity of swept-aerofoil flows to small amplitude wall roughness using resolvent analysis based on wall displacement}

\author[Euryale Kitzinger, Denis Sipp, Olivier Marquet and Estelle Piot]{Euryale Kitzinger$^{1}$%{{\href{https://orcid.org/0000-0001-8419-0929}{\includegraphics{orcid_logo}}}}
, Denis Sipp$^{1\ast}$, Olivier Marquet$^{1}$, and Estelle Piot$^{2}$ }

%\author[2]{Juan G. Santiago}

%\authormark{Rajiv Bharadwaj and Juan G. Santiago}

\address[1]{DAAA, ONERA, University of Paris-Saclay, F-92190 Meudon, France}
\address[2]{DMPE, ONERA, University of Toulouse, F-31055 Toulouse, France}

\corres{*}{Corresponding author. E-mail:
\emaillink{denis.sipp@onera.fr}}


\date{\textbf{Received:} XX 2020; \textbf{Revised:} XX XX 2020; \textbf{Accepted:} XX XX 2020}

\abstract{The receptivity of a laminar boundary layer flow to small amplitude wall roughness is investigated on an ONERA-D swept aerofoil by introducing a dedicated resolvent operator based on linearised small amplitude wall displacements. The singular value decomposition of this operator for a given spanwise wavenumber provides optimal wall roughness and flow responses that maximise an input-output gain.   
%sing the singular value decomposition of the global resolvent operator. %A method to compute the perturbation triggered by a given wall roughness from the singular value decomposition of the resolvent operator is presented.
%This decomposition has the major advantage, when there are dominant singular values, to give a priori informations on the response of the flow to given roughness and also to provide a low-rank approximation of the resolvent operator. This approximation then makes it possible to calculate the perturbation triggered by a given roughness at a reduced numerical cost compared to the direct solution of the linearised Navier-Stokes equations.
%For this purpose, we rely on the preliminary computation of the ``optimal" wall roughness that leads to the most energetic perturbations for various spanwise wavenumber.
At the most receptive spanwise wavenumber, the optimal response is a cross-flow mode associated with an optimal roughness located close to the attachment-line and presenting a wavy shape with a wavevector nearly orthogonal to the external streamlines. The method therefore allows direct identification of the location and structure (chordwise and spanwise wavenumbers) of the most receptive roughness. 
%With the increase of the spanwise wavenumber, the positions of the maximum magnitude of the optimal roughness and response move downstream and upstream respectively.
For various given wall roughness shapes and locations (periodic or compact in the chordwise and/or spanwise directions), an approximation of the response based on the dominant optimal response is shown to accurately match the total response downstream of the roughness. The method therefore allows a straightforward computation of the response of the flow to any given small amplitude roughness.

}

\maketitle

\begin{boxtext}

\textbf{\mathversion{bold}Impact Statement}

The accurate prediction of laminar / turbulent transition on aircraft wings is of major importance for the design of future low-consumption aircraft.
Low atmospheric turbulence rates favour transition triggered by unavoidable surface imperfections. Accurate and straightforward prediction of the flow response to a given wall roughness would pave the way to future shape optimisation or control strategies to efficiently delay transition. However, a general method that accounts for all possible instability mechanisms without any approximation (curvature and non-parallelism) is still lacking. This paper describes a new way to determine the properties of the most critical roughness and its associated response, and to calculate, at a reduced cost, the perturbation amplitude triggered by any given low-amplitude wall roughness.

\end{boxtext}

\section{Introduction}

The laminar or turbulent nature of the boundary layer has a strong impact on aerodynamic aircraft performance. Thus, predicting laminar-turbulent transition and understanding the transitional mechanisms are crucial. The first stage of the laminar-turbulent transition process is receptivity, by which free-stream fluctuations or surface irregularities are transformed into hydrodynamic instabilities within the boundary layer.
These perturbations then grow in the streamwise direction taking advantage of the local instability mechanisms. In the case of swept wings for example, the three main types of instabilities are Attachment-line (AL), cross-flow (CF) and Tollmien-Schlichting (TS) instabilities \citep{reed1989stability}.
Once sufficiently high perturbation amplitudes are reached, nonlinear effects appear. They trigger saturation or new (secondary) instabilities \citep{herbert1988secondary} which may finally lead to transition. When the free-stream perturbations or the surface irregularities are of small amplitude, a linear theory can be used to describe both the initial receptivity process and the first growth phase.   

\citet{deyhle_bippes_1996} have experimentally shown that transverse travelling CF waves are dominant in boundary layers subjected to a high external turbulence rate ($Tu>0.2\%$), while
steady CF waves initiated by surface irregularities dominate for lower turbulence rates, as observed in flight conditions \citep{Carpenter}[$Tu=0.05\%$]. \citet{muller1989experimental} have shown, in the case of a swept flat plate, that steady vortices were triggered by surface irregularities, such as wall roughness.
Numerous studies on the receptivity to surface roughness have been carried out during the last decades.
\citet{Radeztsky1999} studied experimentally the influence of wall roughness in the case of a swept wing where the initial stability characteristics are cross-flow dominated. They noticed that transition was triggered most upstream when the roughness was located near the attachment-line. 
Other experimental studies have been conducted and a review of these works on the influence of surface roughness can be found in \citet{Saric2003}. 
From a numerical point of view, \citet{Crouch1993Receptivity} studied, in a Falkner-Skan-Cooke boundary layer (FSC) within the parallel flow assumption, the competition between receptivity triggered by localised perturbations of the surface geometry and by acoustic waves in the free-stream. He confirmed that at high sweep angle, when considering acoustic waves representative of flight conditions, the boundary layer was dominated by the development of steady cross-flows.
\citet{collis1999receptivity} dealt with the receptivity to spanwise-periodic surface roughness on a swept parabolic cylinder. They showed that surface curvature and, more significantly, non-parallelism plays an important role in receptivity computations.
\citet{schrader2009receptivity} considered non-parallelism for the study of the receptivity to wall roughness of a three-dimensional boundary layer. They modelled a swept wing leading edge with an FSC-like boundary layer with a favourable pressure gradient. They confirmed that, when the turbulence rate was low, steady CF waves dominated the receptivity process over unsteady CF waves induced by free-stream disturbances. They observed that the receptivity mechanism was fully linear when the height of the roughness was below $5\%$ of the displacement thickness.
Comparing the results obtained with a meshed roughness in a DNS and the ones from a parabolised stability equations (PSE) approach with a linear roughness model,
\citet{tempelmann2012swept2} concluded that the linear model was valid up to a roughness height of $~0.1$ of the displacement thickness.

In boundary layer flow, streamwise energy amplification may also occur due to non-modal (local) effects (even when flows are exponentially spatially stable). Contrary to two-dimensional boundary layers, few studies exist on the non-modal growth of three-dimensional boundary layers.
\citet{corbett2001optimal} have computed, in the case of the FSC within parallel-flow assumption, the perturbations with the largest transient gain over a fixed period of time. They showed that, in contrast to the two-dimensional case, both modal and non-modal growths exhibit similar structures, with the optimal perturbations evolving into vortices almost aligned with the direction of the outer flow, which finally trigger  streaks with the lift-up mechanism. These observations were confirmed by \citet{tempelmann2010spatial} who studied the same flow using a spatial framework and PSE allowing them to take into account non-parallel effects. %They also noticed that the perturbation must be introduced near the branch I to excite the steady cross-flow optimally in the case of an accelerated boundary layer.
%\textcolor{green}{ C'est pour amener l'intérêt de faire une étude globale, qui permet de prendre en compte la croissance non-modale mais est-ce que ça vaut le coup vu que je ne vais pas trop développer à ce sujet dans l'article ?}

In order to capture all instability mechanisms (AL, CF, TS, non-modal) and the effects of non-parallelism and surface curvature, global stability analyses (at least in the chordwise direction) on a swept profile were initiated by \citet{tempelmann2012swept1}. They used solutions of the adjoint linearised Navier-Stokes (ALNS) equations to explore the receptivity of cross-flow perturbations on a swept wing of infinite span.
They studied in particular the roughness shape leading to the perturbations with the largest amplitude at the domain outflow. It corresponds to a wavy shape in the chordwise direction that is maximal in the vicinity and just downstream of the attachment-line.
%They confirmed that, in their case, the boundary layer is more receptive to wall roughness than to freestream turbulence. 
The method relies on global direct (resp. adjoint) stability computations with an upstream (resp. downstream) Dirichlet boundary condition determined with a local spatial direct (resp. adjoint) stability computation. The computational domain therefore needs to be chosen in such a way that the same local instability branch be identified at the upstream (direct) and downstream (adjoint) boundaries, which requires some tuning and a precise knowledge of the instability mechanisms.
\citet{thomas2017predicting} used %the same method to include compressibility and surface curvature effects. 
a similar method and a large number of receptivity calculations have been performed for different forcing by computing only once the solutions of the LNS and ALNS equations. In particular, they have confirmed that cross-flow instabilities were more affected by roughness near the attachment-line.

The present paper aims at studying the receptivity to small amplitude wall roughness by introducing a dedicated resolvent analysis \citep{trefethen1993hydrodynamic,book}. Its input is a small amplitude wall deformation (instead of a volume forcing) while the output is, as before, the energy of the perturbation integrated over the volume. The modelling of the wall deformation is based on a linear model, the validity of which has been examined notably by \cite{schrader2009receptivity,tempelmann2012swept2}.
%a method to compute the triggered perturbation at a reduced cost. %This allows to include the surface curvature and to capture the whole spatial structure of the modes. 
%This method is based on the calculation of optimal responses and forcing related to the singular value decomposition (SVD) of the global resolvent operator around the baseflow.
The singular value decomposition of resolvent operators has been widely used to study the energy amplification due to both modal and non-modal mechanisms in transitional flows \citep{sipp2013characterization,symon2018non}. 
It allows the consideration of non-modal mechanisms and to take into account non-parallelism in boundary layer or free-shear flows. Compared to the method used by \cite{thomas2017predicting,tempelmann2012swept1}, it also has the benefit of not relying on relevant inlet / outlet boundary conditions, that come from local direct and adjoint spatial stability solutions: such a method in particular requires  to identify a local spatial modal at the inlet of the computational domain that is connected to a local spatial mode at the outlet of the domain. Hence, only local modal spatial instabilities can be dealt with that method and the inlet and outlet boundaries need to be located in specific regions (for the case considered in the present article, we will even show that it is not possible to follow this mode from the inlet to the outlet of the domain).
Note that a PSE-based method instead of local spatial stability analysis could mitigate this problem: yet, this method still needs to be initialised at the inlet by a local spatial mode and it does not handle local non-modal instabilities \citep{towne2019critical}. In contrast, the present resolvent based method does not suffer from these limitations since it is the optimisation process of the input-output dynamics that automatically identifies the most energetic forcings and responses, even in the case of local non-modal instabilities and in the case of strong non-parallelism.
Also, the use of the resolvent operator to solve the LNS equations gives access to the response amplitude in the vicinity of the location of the roughness and the use of the singular value decomposition has the advantage of providing two orthonormal bases, one for the input space and one for the output space. 
When a singular value is strongly dominant, the analysis gives intrinsic information about the physics of the flow by showing the leading instability mechanism, both in response and in forcing. In this case, once few optimal forcings and responses have been computed, the calculation of the response to a given roughness can be approximated by only computing few scalar products, which results in a reduced computational cost compared to the systematic resolution of the LNS equations. Moreover, the prediction of the full response based on the dominant optimal forcings / responses will be accurate downstream of the roughness, which is usually sufficient for laminar / turbulent transition predictions.
We will assess the accuracy of the prediction when using the dominant singular value for both periodic and compact roughness in the chordwise and spanwise directions. 
The method will be illustrated on the swept ONERA-D aerofoil, a configuration that was already studied (base-flow and neutral stability curves of attachement-line, cross-flow and Tollmien-Schlichting perturbations) in \citet{kitzinger_leclercq_marquet_piot_sipp_2023}. 

The outline of the paper is the following.
In \S $2$, we describe the flow configuration, in \S $3$ the wall displacement based resolvent analysis and the approximations and in \S $4$ the numerical methods. In \S $5$, the results are illustrated for the ONERA-D aerofoil. The properties of the optimal roughness and responses are explored in \S $5.1$ and the perturbations triggered by particular roughness are studied in \S $5.2$. We validate in particular in this last sub-section the low-rank approximation against the exact response (given by the resolvent) for various roughness shapes.

%\textcolor{brown}{Est-ce que l'un de vous peut corriger et/ou modifier l'introduction et notamment le paragraphe sur la comparaison de l'analyse de résolvent avec les méthodes précédemment utilisées ? C'est le coeur des critiques des reviewers et je pense que vous arriverez à être plus pertinents que moi. Il faudrait modifier les réponses à la remarque $(iv)$ du reviewer 1 et $(i)$ du reviewer 3 en conséquence.}

\section{Flow configuration}

\begin{figure}
\vspace{0.5cm}
\begin{center}
\begin{tabular}{rlrl}
(a) & & (b) &\\
& \hspace{-1cm}\includegraphics[trim = 40mm 55mm 125mm 50mm, clip,width=0.6\textwidth]{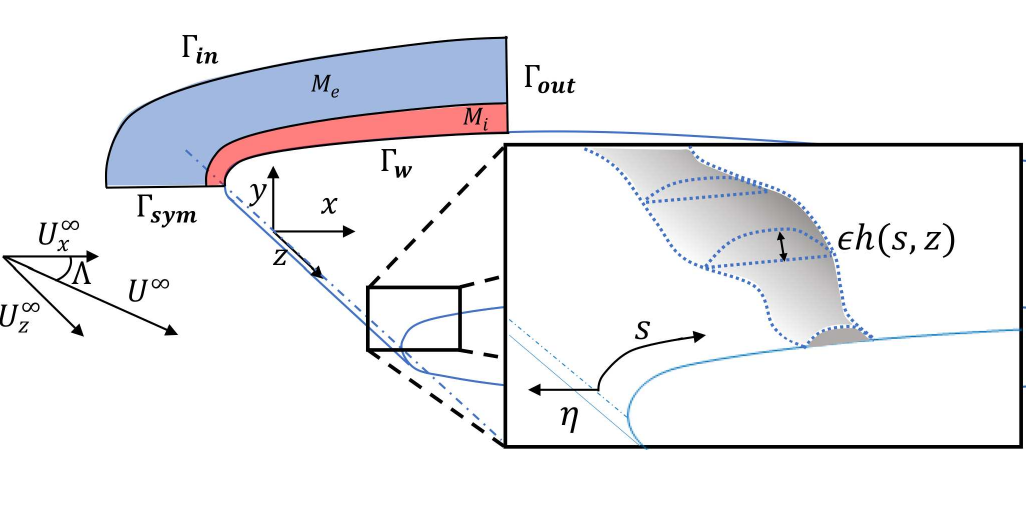} & & \includegraphics[trim = 110mm 80mm 120mm 30mm, clip,width=0.4\textwidth]{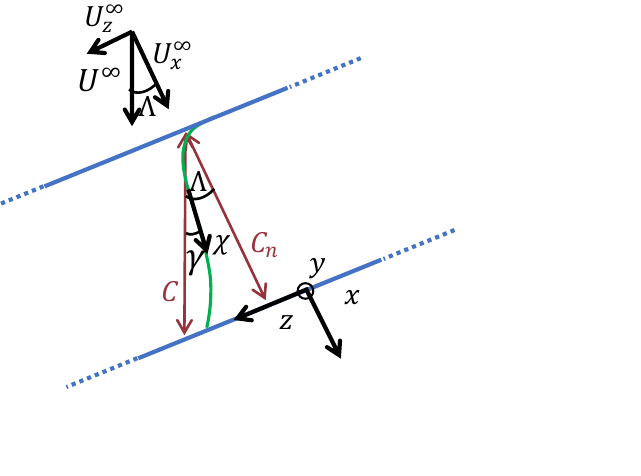}
\end{tabular}
\end{center}
\caption{$(a)$: Schematic of the mesh and flow configuration with a zoom on a roughness of any shape. $(b)$: Angles and coordinate systems are indicated. An external streamline is illustrated in green. The blue lines correspond to the leading/trailing edges. }
\label{fig:configuration}
\end{figure}

We investigate the incompressible flow around an {aerofoil} of chord $C$ and of infinite span. The origin of the orthonormal coordinate system $(x,y,z)$ is located at the leading edge of the aerofoil, whose direction is denoted $z$. As shown in figure \ref{fig:configuration}, the $x$- and $y$-directions are orthogonal to the leading edge and to the symmetry plane of the aerofoil, respectively. The uniform upstream velocity is of constant magnitude $U^{\infty}$ and oriented in a direction defined by the sweep angle $\Lambda$ and the angle of attack $\alpha$, that is zero in the present study. The sweep velocity denoted $U_z^\infty=U^\infty \sin \Lambda$ is the component of the upstream velocity in the spanwise direction while the chordwise velocity $U_x^\infty=U^\infty \cos \Lambda$ is the component in the chordwise direction. Note that $C$ denotes the aerofoil's chord in the direction of the upstream velocity while $C_n= C \cos \Lambda $ is the chord normal to the aerofoil's leading edge. In the following, all variables are made non-dimensional using the chord $C_n$ and the velocity $U_x^\infty$. The incompressible flow is entirely characterised by two non-dimensional parameters: the sweep angle $\Lambda$ and the Reynolds number  
$Re_{C_n}=U_x^\infty C_n / \nu$, $\nu$ being the kinematic viscosity of the fluid.  
The local coordinate system $(s,\eta,z)$
is shown in the close-up view of figure \ref{fig:configuration}(a). $s$ is the curvilinear abscissa along the surface of the profile in the plane orthogonal to the spanwise $z$-direction and $ \eta $ the local wall normal direction. Thus, $\eta=0 $ corresponds to the aerofoil's wall. The local direction of the streamline of the external baseflow velocity field $\chi$ is shown in figure \ref{fig:configuration}(b).

%%%%%%%%%%%%%%%%%%%%%%%%%%%%%%%%%%%%%%%%%%%%%%%%
%%%%%%%%%%%%%%%%%%%%%%%%%%%%%%%%%%%%%%%%%%%%%%%%%
\section{Theory}\label{part:equations}
We investigate the incompressible three-dimensional steady flow surrounding a rough aerofoil infinitely long in the spanwise direction $z$. The non-dimensional flow velocity $\boldsymbol{U}_{T}(x,y,z)$ and pressure $P_{T}(x,y,z)$ fields are governed by the steady incompressible Navier-Stokes equations  
\begin{eqnarray}\label{eqn:NS3D}
\left( \boldsymbol{U}_{T} \cdot \boldsymbol{\nabla}  \right) \boldsymbol{U}_{T} + \boldsymbol{\nabla} P_{T} - Re_{C_{n}}^{-1}\boldsymbol{\nabla}^2 \boldsymbol{U}_{T} = \boldsymbol{0}, \;\; \boldsymbol{\nabla} \cdot \boldsymbol{U}_{T} = 0  \mbox{ in } \Omega_{r}, \;\;\;\; \boldsymbol{U}_{T} = \boldsymbol{0} \mbox{ on } \Gamma_{r}, 
\end{eqnarray}
where the gradient is classicaly defined as $\boldsymbol{\nabla}=(\partial_x,\partial_y,\partial_z)^T$. The momentum and mass equations are satisfied in the spatial domain $\Omega_{r}$ surrounding the aerofoil and the roughness. The boundary formed by the aerofoil and the roughness is denoted $\Gamma_r$, on which a no-slip velocity condition is imposed. \\

As sketched in the close-up view of figure 
\ref{fig:configuration}(a), we further assume that the height of any wall roughness, denoted $\epsilon h(s,z)$, is infinitesimally small compared to the thickness / chord of the smooth aerofoil, the surface of the latter being denoted $\Gamma_w$. In the local curvilinear coordinate system, the surface of $ \Gamma_r$ is thus described by $(s,\eta=\epsilon h(s,z), z)$ with $\epsilon \ll 1 $, while the surface of the smooth aerofoil $ \Gamma_w $ is $(s,\eta=0, z)$. The roughness being considered as a small perturbation of the smooth aerofoil geometry, we decompose the flow variables as 
\begin{equation*}
(\boldsymbol{U}_{T},P_{T})=
(\boldsymbol{U},P)+ \epsilon (\boldsymbol{u},p),
\end{equation*}
where $(\boldsymbol{U},P)$ denotes the flow over the smooth aerofoil while $(\boldsymbol{u},p)$ is the flow perturbation induced by the roughness. This decomposition is injected into the governing equations \eqref{eqn:NS3D} to obtain, at zeroth-order the equations the flow around the smooth aerofoil, and at $\epsilon$ order the wall roughness perturbation equations. 
In particular, the no-slip boundary condition along $ \Gamma_r$ reads 
\begin{equation*}
    \boldsymbol{0}=\boldsymbol{U}_{T}\left(s,\epsilon h(s,z),z\right) \approx \boldsymbol{U}\left(s,0,z\right)+\epsilon \left[ \boldsymbol{u} + h \partial_\eta \boldsymbol{U}
    %\left(\boldsymbol{n} \cdot\boldsymbol{\nabla}\right)\boldsymbol{U}
    \right](s,0,z) +\cdots.
\end{equation*}
%where $\textbf{n}$ denotes the unit vector normal to the surface of the smooth aerofoil. 
This Taylor expansion shows that, at first order in $\epsilon$, the small roughness is equivalent to the velocity perturbation over the smooth aerofoil  
\begin{equation}\label{noslip}
\boldsymbol{u}(s,0,z) = - h \partial_\eta \boldsymbol{U}.%\left(\boldsymbol{n} \cdot\boldsymbol{\nabla}\right)\boldsymbol{U} 
\end{equation}
The validity of this linearised velocity condition for modelling the effect of a roughness on a boundary layer flow was addressed by \citep{schrader2009receptivity,tempelmann2012swept2} who found that this is a valid approximation for roughness height below $0.05\delta^*$, $\delta^*$ being the displacement thickness of the boundary layer.

\begin{comment}
In our study, we consider a homogeneous configuration in the spanwise direction with a steady 2.5D baseflow $\boldsymbol{Q}(x,y)=(U_x,U_y,U_z,P)(x,y)=(\boldsymbol{U},P)(x,y)$ and a steady 3D perturbation  $\epsilon\boldsymbol{q}(x,y,z)=\epsilon(u_x,u_y,u_z,p)(x,y)=\epsilon(\boldsymbol{u},p)(x,y,z)$ triggered by a small amplitude ($\epsilon\ll1$) steady wall normal displacement $\eta=\epsilon h(s,z)$. The total flow $\boldsymbol{Q}_{tot}=\boldsymbol{Q}(x,y)+\epsilon\boldsymbol{q}(x,y,z)$ is governed by the steady Navier-Stokes equations % (noted $\mathcal{N}$):
%\begin{equation}\label{NavierStokes}
$    \mathcal{N}(\boldsymbol{Q}_{tot},Re_{C_n}, \Lambda)=\boldsymbol{0}$
%\end{equation}
and no-slip boundary conditions at the wall $ \Gamma_w $ expressed as:
\begin{equation}\label{noslip}
    \boldsymbol{U}_{tot}\left(s,\epsilon h(s,z),z\right)=\boldsymbol{0}\Rightarrow \boldsymbol{U}\left(s,0\right)+\epsilon\left(\boldsymbol{u}+h\left(\boldsymbol{n} \cdot\boldsymbol{\nabla}\right)\boldsymbol{U}\right)+\cdots=\boldsymbol{0}.
\end{equation}
where $\textbf{n}$ denotes the wall normal vector. The last equation of \eqref{noslip} comes from the first-order Taylor series expansion.
\end{comment}

%Throughout the paper, we will use the coordinate system $(s,\eta,z)$ or $(s,z)$ when referring to wall variables, and the system $(x,y,z)$ for variables in the domain $\Omega$.

\subsection{Steady flow over the smooth aerofoil}
The equations governing the baseflow velocity field $\boldsymbol{U}(x,y)=(U,V,W)^{T}$ around the swept smooth aerofoil are the steady Navier-Stokes equations:
\begin{eqnarray}\label{eqn:NS2D5}
\left( \boldsymbol{U} \cdot \boldsymbol{\nabla}  \right) \boldsymbol{U} + \boldsymbol{\nabla} P - Re_{C_{n}}^{-1} \boldsymbol{\nabla}^2  \boldsymbol{U} = \boldsymbol{0}, \;\; \boldsymbol{\nabla} \cdot \boldsymbol{U}= \boldsymbol{0}  \mbox{ in } \Omega, \;\;\;\; \boldsymbol{U} = \boldsymbol{0} \mbox{ on } \Gamma_w.
\end{eqnarray}
%where the two-dimensional gradient and Laplacian operators are defined as $\boldsymbol{\nabla}_{0}=(\partial_x,\partial_y,0)^{T}$, $\boldsymbol{\Delta}_{0}=(\Delta_0, \Delta_0 ,\Delta_0 )^{T}$ and $\Delta_{0}=(\partial_{xx},\partial_{yy},0)^{T}$.
As the aerofoil is symmetric and the angle of attack is zero, the domain $\Omega$ may be restricted to the upper half domain, i.e. $y>0$. A symmetric boundary condition $(\partial_y U_x, U_y, \partial_y U_z)=(0,0,0)$ is applied at the symmetric boundary $\Gamma_{sym}=\{x<0,y=0\}$. As shown in figure \ref{fig:configuration}(a), the spatial domain $\Omega$ does not extend up to the trailing-edge of the aerofoil. To specify the boundary conditions at the inflow $\Gamma_{in}$ and outflow $\Gamma_{out}$ boundaries of this domain, we first determine the symmetric potential velocity field $(U_x^p,U_y^p,U_z^p)=(\partial_y \psi, -\partial_x \psi,U_z^\infty=\tan(\Lambda))$ with the stream-function $\psi$ satisfying the Laplace equation $\Delta \psi = 0$, in a sufficiently large domain $\Omega_{p}$ so that we may impose the uniform velocity condition $\psi = y$ on the far-field boundary $ \Gamma_p $. On $ \Gamma_w$ and $ \Gamma_{sym}$, we set $\psi=0$, which allows the smooth aerofoil to be a streamline and the flow field to be symmetric. This potential velocity is then imposed as a Dirichlet boundary condition at the inlet boundary $\Gamma_{in}$ of the domain $\Omega$, i.e. 
\begin{equation}
    (U_x,U_y,U_z)=\left(U_x^{p},U_y^{p},U_z^\infty\right) \textrm{ on } \Gamma_{in},
\end{equation}
while the pressure field of the potential flow $P^p=[1-(U_x^{p})^2-(U_y^{p})^2]/2$ and a Neumann condition for the velocity field are applied at the outlet:
\begin{equation}
\partial_x \textbf{U} =\textbf{0} \mbox{ and } P=P^{p} \mbox{ on } \Gamma_{out}.
\end{equation}

\subsection{Wall-displacement induced perturbation}\label{par:S}
We consider a roughness that may be harmonic, periodic or compact in either the curvilinear $ s $- or spanwise $ z $-directions.
The linear equations governing the three-dimensional flow perturbation $(\boldsymbol{u},p)(x,y,z)$ induced by this roughness are:
\begin{eqnarray}
\label{eqn:Pert2D5}
\left( \boldsymbol{u} \cdot \boldsymbol{\nabla}  \right) \boldsymbol{U} + \left( \boldsymbol{U} \cdot \boldsymbol{\nabla}  \right) \boldsymbol{u}  + \boldsymbol{\nabla} p - {Re_{C_{n}}^{-1}} \boldsymbol{\nabla}^2 \boldsymbol{u} = \boldsymbol{0}, \;\; \boldsymbol{\nabla} \cdot \boldsymbol{u}= 0 \mbox{ in } \Omega  \;\; \;\; \;\;\;\;\;\;
\boldsymbol{u} = - h \partial_\eta \boldsymbol{U} \mbox{ on } \Gamma_w.
\end{eqnarray}
As explained before, the presence of the roughness over the smooth aerofoil is accounted for with the linearised wall boundary condition. The solution of the above linear problem with non-homogeneous boundary conditions at the wall is the sum of the solution of the problem with homogeneous condition and a particular forced solution. We choose the Reynolds number and sweep angle so as to have a globally stable base-flow, so that all homogeneous solutions tend to zero at large times. We hence only consider the forced response. Interestingly, the spanwise-dependency of the flow perturbation is only due to the spanwise dependency of the roughness $h(s,z)$, since the velocity $\boldsymbol{U}$ is invariant in the spanwise direction. Hence, any roughness and triggered perturbation can be further decomposed into Fourier modes as 
\begin{eqnarray}\label{eqn:Fourier}
    h(s,z) = \int_{-\infty}^{\infty} \hat{h}_{\beta}(s) e^{i\beta z} \, d \beta,\;\;
    \boldsymbol{u}(x,y,z) = \int_{-\infty}^{\infty} \boldsymbol{\hat{u}}_{\beta}(x,y) e^{i\beta z}  \, d \beta
\end{eqnarray}
where $\beta$ is the (real) spanwise wavenumber and the corresponding (complex) Fourier modes of the wall roughness and flow response are denoted 
$\hat{h}_{\beta}(s)$ and $\boldsymbol{\hat{u}}_{\beta}(x,y)$, respectively.
Note that the case of periodic roughness in $z$ may be handled by stating that $ \hat{h}_\beta(s)=\sum_m \delta(\beta-m\beta_0) \hat{h}_{m\beta_0}(s)$ with $ \beta_0=2\pi/L_z$ being the fundamental wavenumber in $z$ and $ \delta(\beta) $ the Dirac function at $ \beta=0$.
%The same decomposition holds for the pressure field. 
Since the three-dimensional roughness $h(s,z)$ and flow perturbation $\boldsymbol{u}(x,y,z)$ are real quantities, the corresponding complex Fourier modes satisfy $\hat{h}_{-\beta} = \overline{\hat{h}_{\beta}}$ and $\boldsymbol{\hat{u}}_{-\beta} = \overline{\boldsymbol{\hat{u}}_{\beta}}$ where $\overline{\left(\cdot\right)}$ denotes the complex conjugate.

\begin{comment}
The above Fourier decomposition can thus be reformulated as
\begin{eqnarray}\label{eqn:FourierReal}
    h(s,z) &=& \int_{0}^{\infty} \left( \hat{h}_{\beta}(s) e^{i\beta z} + \overline{\hat{h}_{\beta}}(s) e^{-i\beta z}  \right) d \beta \\
    \boldsymbol{u}(x,y,z) &=& \int_{0}^{\infty} \left( \boldsymbol{\hat{u}}_{\beta}(x,y) e^{i\beta z} + \overline{\boldsymbol{\hat{u}}_{\beta}}(x,y) e^{-i\beta z} \right) d \beta \nonumber
\end{eqnarray}
\end{comment}
Injecting this Fourier decomposition into \eqref{eqn:Pert2D5}, we obtain 
\begin{equation}
\label{eqn:Fourier2D5}
\left\{
\begin{array}{lr}
\left( \boldsymbol{\hat{u}}_{\beta} \cdot \boldsymbol{\nabla}  \right) \boldsymbol{U} + \left( \boldsymbol{U} \cdot \boldsymbol{\nabla}_{\beta}  \right) \boldsymbol{\hat{u}}_{\beta}  + \boldsymbol{\nabla}_{\beta} \hat{p}_{\beta} - {Re_{C_{n}}^{-1}} \boldsymbol{\nabla}_{\beta}^2  \boldsymbol{\hat{u}}_{\beta} = \boldsymbol{0}, \;\; \boldsymbol{\nabla}_{\beta} \cdot \boldsymbol{\hat{u}}_{\beta} = 0 \mbox{ in } \Omega \; \Rightarrow \mathcal{L}_\beta\cdot\left(\boldsymbol{\hat{u}}_{\beta},\hat{p}_{\beta}\right)^T=\boldsymbol{0} \mbox{ in } \Omega , \\
\boldsymbol{\hat{u}}_{\beta} = - \hat{h}_\beta \partial_\eta \boldsymbol{U}\mbox{ on }  \Gamma_w\; \Rightarrow \hat{u}_{\beta}=\mathcal{F}\hat{h}_\beta\mbox{ on }  \Gamma_w,
\end{array}
\right.
\end{equation}
where the gradient and Laplacian operators applied on the Fourier modes are defined as $\boldsymbol{\nabla}_{\beta}=(\partial_x,\partial_y,\rm{i} \beta)^{T}$. $\mathcal{L}_\beta$ is the linearised Navier-Stokes operator and $\mathcal{F}$ is the operator that transforms the wall normal displacement $\hat{\boldsymbol{h}}_{\beta}$ into a velocity $\hat{\boldsymbol{u}}_{\beta}$ with appropriate Dirichlet boundary conditions on the wall and zeros inside the domain. The boundary conditions at the inflow and outflow boundaries are 
$\boldsymbol{\hat{u}}_{\beta} = \textbf{0} $ on $\Gamma_{in} $ and $\hat{p}_{\beta} \boldsymbol{e}_x-Re_{C_n}^{-1} \partial_x \boldsymbol{\hat{u}}_{\beta} = 0$ on $ \Gamma_{out} $
and we restrict our analysis to symmetric perturbations, i.e. $\left( \partial_y \hat{u}_{\beta,x}, \hat{u}_{\beta,y}, \partial_y \hat{u}_{\beta,z} \right) = (0,0,0) \mbox{ on }  \Gamma_{sym}$.
We can thus calculate the flow perturbation triggered by a roughness over the smooth aerofoil without modifying the geometry, by using the steady flow $\boldsymbol{U}$ over the smooth aerofoil and by applying the linearised wall velocity condition at its surface.
After spatial discretisation, the equations \eqref{eqn:Fourier2D5} with the above boundary conditions can be recast in an input-output form:
\begin{equation}\label{ResolventDiscret}
    \boldsymbol{\hat{u}}_{\beta}=R_\beta \, \hat{\boldsymbol{h}}_{\beta}, 
\end{equation}
where $R_{\beta}=P^*L_\beta^{-1}PF$. 
The matrices $L_\beta$ and $ F$ are respectively the discrete form of the continuous operators $\mathcal{L}_\beta$ and $\mathcal{F}$ defined in equations \eqref{eqn:Fourier2D5}.
The matrix $ P $ designates the prolongation operator which adds a zero pressure component to a given velocity vector.
 %Denoting $n_{h,w}$ the number of degrees of freedom of the wall forcing and $n_{u,\Omega}$ the number of degrees of freedom of the velocity vector in the domain, $F$ is of dimension $n_{u,\Omega} \times n_{h,w}$. 

$\hat{\boldsymbol{h}}_{\beta}$ denotes the discrete vector of the function $\hat{h}_{\beta}(s)$, and $\boldsymbol{\hat{u}}_{\beta}$ is the discrete velocity vector of the continuous vectorial field $\boldsymbol{\hat{u}}_{\beta}(x,y)$. In the following, the spatial coordinates $(x,y)$ are specified to distinguish the continuous and discrete forms of the velocity vector. $R_{\beta}$ will be called the "resolvent" matrix  (even though it is not a square matrix) and can be considered as a transfer function taking 
a wall roughness Fourier mode $\boldsymbol{\hat{h}}_{\beta}$ as input 
and giving the flow-perturbation Fourier mode $\boldsymbol{\hat{u}}_{\beta}$ as output.

\begin{comment}
where $R_{\beta}=P^*L_\beta^{-1}PF$ and $F$ the discretised version of the wall boundary condition \eqref{CL_petites_perturbations_wall_resolvent}, that transforms the wall normal displacement vector $\boldsymbol{\hat{h}}$ into a velocity vector $\boldsymbol{\hat{u}}$ containing appropriate Dirichlet boundary conditions at the wall and zero values inside the domain. Denoting $n_{h,w}$ the number of degrees of freedom of the wall forcing and $n_{u,\Omega}$ the number of degrees of freedom of the velocity vector in the domain, $F$ is of dimension $n_{u,\Omega} \times n_{h,w}$. 
$L_\beta$ and $P$ are respectively the discrete form of $\mathcal{L}_\beta$ and $\mathcal{P}$ where the latter is the prolongation operator which adds a zero pressure component to a given velocity vector. Matrix $R_\beta$ can be considered as a transfer function taking a wall roughness as input and giving the triggered perturbation as output.
\end{comment}

\subsection{Resolvent analysis with wall displacements}\label{par:Resolvent}
We will now determine two orthonormal bases, one for the (input) wall roughness Fourier modes and one for the (output) flow-response Fourier modes. For this, we define the kinetic energy of the trigerred perturbation $\langle\boldsymbol{\hat{u}}_{\beta}(x,y),\hat{\boldsymbol{u}}_{\beta}(x,y)\rangle = \int_\Omega\boldsymbol{\hat{u}}_{\beta}^*(x,y) \boldsymbol{\hat{u}}_{\beta}(x,y) \; dxdy $ in the domain $\Omega$ as a measure of the output space, where $(\cdot)^*$ refers to the transconjugate and
$\langle{\hat{h}}_{\beta}(s),{\hat{h}}_{\beta}(s)\rangle_{w} = 
\int_{\Gamma_w} \overline{\hat{h}_{\beta}(s)}\hat{h}_{\beta}(s)\;ds$
as a measure of the input space. The discrete form for each measure is then denoted $\boldsymbol{\hat{u}}_{\beta}^* Q_u \boldsymbol{\hat{u}}_{\beta}$ and $ \boldsymbol{\hat{h}}_{\beta}^* Q_h \boldsymbol{\hat{h}}_{\beta}$, respectively. Then, we introduce the energetic gain $G_{\beta}$ to be maximised as the ratio between the output and input measures, i.e.
\begin{equation}
    G_{\beta}=\frac{\boldsymbol{\hat{u}}_{\beta}^* Q_u \boldsymbol{\hat{u}}_{\beta}}{\boldsymbol{\hat{h}}_{\beta}^* Q_h \boldsymbol{\hat{h}}_{\beta}}=\frac{\boldsymbol{\hat{h}}_{\beta}^* R_\beta^* Q_u R_\beta \boldsymbol{\hat{h}}_{\beta}}{\boldsymbol{\hat{h}}_{\beta}^* Q_h \boldsymbol{\hat{h}}_{\beta}},
\end{equation}
where the input-output relation \eqref{ResolventDiscret} has been used in the numerator.
%with the input and output related by \eqref{ResolventDiscret} and the following scalar products:
%\begin{eqnarray}
%\langle\boldsymbol{u}_1(x,y),\boldsymbol{u}_2(x,y)\rangle&=&\int_\Omega\boldsymbol{u}_1(x,y)^*\boldsymbol{u}_2(x,y) \; dxdy=\boldsymbol{u}_1^* Q_u \boldsymbol{u}_2, \\
%\langle{h}_1(s),{h}_2(s)\rangle_w&=&\int_{\Gamma_w} \overline{h_1(s)}h_2(s) \; %ds=\boldsymbol{h}_1^*Q_h \boldsymbol{h}_2,
%\end{eqnarray}
%we have:
%\begin{equation}
%    G(\boldsymbol{\hat{h}})=\frac{\boldsymbol{\hat{u}}^* Q_u \boldsymbol{\hat{u}}}{\boldsymbol{\hat{h}}^* Q_h \boldsymbol{\hat{h}}}=\frac{\boldsymbol{\hat{h}}^* R_\beta^* Q_u R_\beta\boldsymbol{\hat{h}}}{\boldsymbol{\hat{h}}^* Q_h \boldsymbol{\hat{h}}},
%\end{equation}
The solution of the optimisation problem $\max_{\boldsymbol{\hat{h}}_{\beta}}G_{\beta}$ and the optimal roughness / responses are finally obtained by solving the two problems:
\begin{equation}\label{eq:Eigen}
    R_\beta^* Q_u R_\beta\, \boldsymbol{\hat{h}}_{\beta,j} =\sigma_{\beta,j}^2 Q_h \, \boldsymbol{\hat{h}}_{\beta,j}, \;\;\;\; \boldsymbol{\hat{u}}_{\beta,j}=\sigma_{\beta,j}^{-1} R_\beta \, \boldsymbol{\hat{h}}_{\beta,j}. 
\end{equation}
The optimal roughness $\boldsymbol{\hat{h}}_{\beta,j}$ and optimal flow responses
are normalised as $\langle{\hat{h}}_{\beta,j}(s),{\hat{h}}_{\beta,j}(s)\rangle_{w}=1 $ and $\langle\boldsymbol{\hat{u}}_{\beta,j}(x,y),\hat{\boldsymbol{u}}_{\beta,j}(x,y)\rangle=1$. 
 Note that a fully continuous framework for the definition of these quantities also exists but is not shown here for conciseness. The set of eigenvectors $({\hat{h}}_{\beta,j}(s))_{j\geq1}$ form an orthonormal basis of the forcing space with respect to $ \langle \cdot,\cdot \rangle_w $, while the optimal responses $(\boldsymbol{\hat{u}}_{\beta,j})(x,y)_{j\geq1}$ constitute an orthonormal basis of the response space with respect to $ \langle \cdot,\cdot \rangle $. %$(\boldsymbol{\hat{u}}_{\beta,j})_{j\geq1}$, $(\boldsymbol{\hat{h}}_{\beta,j})_{j\geq1}$ and $(\sigma_{\beta,j})_{j\geq1}$ correspond to the left-singular vectors, right-singular vectors and singular values of $R_{\beta}$, respectively.
In our study, we sort the singular values such that $\sigma_{\beta,1}\geq\sigma_{\beta,2}\geq\sigma_{\beta,3}\geq...$ so that the optimal forcing ${\hat{h}}_{\beta,1}(s)$ is related to $\sigma_{\beta,1}^2=\max_{\boldsymbol{\hat{h}}}G_\beta$.

We can now compute the response $\boldsymbol{\hat{u}}_{\beta}(x,y)$ in equation \eqref{ResolventDiscret} using the resolvent mode bases:
\begin{equation}\label{eq:ProjBasis}
    \boldsymbol{\hat{u}}_{\beta}(x,y)=\sum_{j\geq1} \sigma_{\beta,j} \; \gamma_{\beta,j}  \; \boldsymbol{\hat{u}}_{\beta,j}(x,y),  
\end{equation}
with $\gamma_{\beta,j}=\langle \hat{h}_{\beta,j}(s),\hat{h}_{\beta}(s)\rangle_w$ the projection coefficients of the wall roughness Fourier mode $\hat{h}_{\beta}(s)$ onto the resolvent modes $\hat{h}_{\beta,j}(s)$. In the case where the first singular value is much larger than the following ones, which is generically the case when an instability mechanism is at play, we can neglect the contribution of the next terms:
if $\sigma_{\beta,1}\;\lvert  \gamma_{\beta,1} \rvert\gg \sigma_{\beta,j} \; \lvert  \gamma_{\beta,j} \rvert \; \mbox{    }\forall {j\geq 2} $,
we can consider only the contribution of the dominant optimal response:
$ \boldsymbol{\hat{u}}_{\beta}(x,y) \approx \sigma_{\beta,1} \; \gamma_{\beta,1}  \;  \boldsymbol{\hat{u}}_{\beta,1}(x,y)$.

%We then understand how the knowledge of the optimal forcing and the associated optimal response is an essential tool to identify the behavior of the system subjected to a forcing.
We then obtain an approximation of the response $ \boldsymbol{u}(x,y,z) $ triggered by the roughness $ h(s,z)$ following% Indeed, once the optimal roughness and response have been computed, we can obtain the triggered perturbation from a simple scalar product.

\begin{equation}\label{eq:LowRankProj} 
\boldsymbol{u}(x,y,z) \approx \int_{-\infty}^{+\infty} \sigma_{\beta,1} \; \gamma_{\beta,1}   \; \boldsymbol{\hat{u}}_{\beta,1}(x,y)\; e^{i\beta z}\;d\beta,
  \end{equation}
and the local perturbation energy averaged over the spanwise direction reads:
%    \begin{equation} \label{eq:mean2} \langle\|\boldsymbol{u}\|^2\rangle_z(x,y)=\int_{-\infty}^{+\infty}  \left[
%    \sum_{j\geq1}\sigma_{\beta,j}}^2|\gamma_{\beta,j}|^2\|\boldsymbol{\hat{u}}_{\beta,j}\|^2(x,y)
%    \right] \; d\beta.
%  \end{equation}
    \begin{equation} \label{eq:mean2} \langle\|\boldsymbol{u}\|^2\rangle_z(x,y)=\int_{-\infty}^{+\infty}  \sigma_{\beta,1}^2\; |\gamma_{\beta,1}|^2 \;\|\boldsymbol{\hat{u}}_{\beta,1}\|^2(x,y)
 \; d\beta,
  \end{equation}
where $ \langle \cdot \rangle_z =\lim_{L_z \rightarrow \infty} (1/L_z) \int_0^{L_z} (\cdot) \;dz$.
Explicit approximations may then be obtained by evaluating the integral with a fourth order extended Simpsons's rule \citep{press2007numerical} over a finite wavenumber range, as discussed in section \ref{results:ellipse}.

%where the subscript $_{j,\beta}$ denotes the $i^{th}$ element of the SVD for the spanwise wavenumber $\beta$.

%As will be seen in the application part of \ref{part:ResponseGivenForcing}, the approximations already mentioned in equation \eqref{LowRank1} can be used in order to reduce the computation of the response to the contribution of only few vectors of $({\hat{h}}_{\beta,j}(s))_{j\geq1}$ and $(\boldsymbol{\hat{u}}_{\beta,j}(x,y))_{j\geq1}$.

%Throughout the paper, when the value of $\beta$ is fixed and unambiguous, the subscript $_\beta$ will be removed to lighten the notations.

\section{Numerical methods}

All numerical aspects are handled with FreeFEM \citep{hecht:hal-01476313}, an open-source partial-differential-equations  solver that allows to implement spatial discretisation with the Finite Element method. Solutions of resulting large-scale linear problems are computed on multiple processors using the FreeFEEM's interface with the Portable, Extensible Toolkit for Scientific Computation (PETSc)
\citep{petsc-user-ref}. The interface with the SLEPc solver \citep{Hernandez:2005:SSF} is used to compute the solution of eigenvalue problems. We refer to  
\cite{moulin2019augmented} for  didactic examples of these two interfaces in the context of linear stability analysis of large-scale hydrodynamic eigenvalue problems. 
Taylor-Hood finite elements are used for the spatial discretisation of non-linear base-flow equations \eqref{eqn:NS2D5} and linear perturbation equations \eqref{eqn:Fourier2D5}. The velocity and pressure fields are respectively expanded on second-order ($P_2$) and first-order ($P_1$) Lagrange finite elements.
A Streamline-Upwind Petrov–Galerkin (SUPG) method together with grad–div stabilisation \citep{ahmed2019numerical} is used to compute the base-flow solution while no SUPG stabilisation is considered for the perturbation. A Newton-Raphson method is implemented in FreeFEM to compute solutions of the nonlinear base-flow equations \eqref{eqn:NS2D5}, using the potential flow solution as initial condition of this iterative algorithm. The direct sparse LU-solver MUMPS \citep{MUMPS:1,MUMPS:2} is called within PETSc to compute the solution of linear problems at each iteration of the algorithm. 
For the resolvent analysis, a Krylov-Schur algorithm is used to compute the largest eigenvalues of the eigenproblem \eqref{eq:Eigen}. The application of the matrices $ R_\beta $ and $ R_\beta^* $ to input vectors provided by the Krylov-Schur algorithm requires the solution of linear problems that are obtained with the MUMPS solver, again.

%\textcolor{brown}{Olivier, est-ce que tu peux modifier le paragraphe précédent sur la méthode numérique ? Notamment par rapport à la méthode de shift-invert qui est seulement là pour corriger une faiblesse de FreeFem si je me souviens bien. Il faut peut-être modifier la réponse au commentaire $(v)$ du reviewer 3 en conséquence.}

As shown in figure \ref{fig:configuration}, we consider a two-dimensional domain $\Omega$ covering the upper half of the aerofoil and which extends up to $15\%$ in the chordwise $x$-direction. Two different body-fitted meshes are used to compute the base-flow and resolvent analysis. For the base-flow solution, this mesh is composed of $120 000$ triangles. It is made up of an internal $M_i$ part (red in figure \ref{fig:configuration}, $100 000$ triangles) and an external $M_e$ part (blue, $20 000$ triangles). For the resolvent analysis, we use only the internal mesh $M_i$, the boundary $ \Gamma_{in}$ being sufficiently far from the profile so that 0-Dirichlet boundary conditions may be imposed. For more details, we refer to \cite{kitzinger_leclercq_marquet_piot_sipp_2023} where the same  mesh was used to perform global stability analysis of the baseflow solution.
In particular, the effect of the chordwise extension of the domain and of grid refinement was assessed for the computation of base-flows and neutral curves (marginal eigenvalues of the Jacobian $ L_\beta $). A similar convergence study (not reported here) was performed for the singular values and resolvent modes to insure that reported results are robust to spatial discretisation.

\section{Results}

In the present study, we consider the parameters 
$Re_{C_n}=1.39 \times 10^6$, $\Lambda=65.8^\circ$,
which corresponds to a globally stable flow for all spanwise wavenumbers \citep{kitzinger_leclercq_marquet_piot_sipp_2023} and thus allows the input-output analysis described in \S \ref{par:Resolvent}. In that study, the Reynolds number and sweep angle were chosen to allow the computation of the neutral curve associated to the various instabilities, which remains simple only at high sweep angles. Yet, for the present study, which relies on the more robust resolvent analysis to characterise the instabilities, lower sweep angles could have been chosen. We decided to keep the initial parameters to focus the paper on the novelty, which is methodological. Hence, extensive discussions about the base-flow solution can be found in \cite{kitzinger_leclercq_marquet_piot_sipp_2023}. We recall that it was validated by comparing the external streamline velocity component and the cross-flow component within the boundary layer with those obtained by using an ONERA in-house boundary layer code which solves the Prandtl's equations \citep{houdeville1992three}. We observed a close agreement between the results obtained with both methods.

For the description of the results, the spanwise wavenumber $\beta$ of the perturbations will be scaled with $\Delta$ which is a measure of the boundary layer thickness at the attachment-line based on the potential flow \citep{Mack2011}. It is defined as
$\Delta = \left. \partial U_s^{p}/\partial s\right|_{x=0,y=0}$,
where $ U_s^{p} $ denotes the $s$-component of the potential flow solution. In \citet{kitzinger_leclercq_marquet_piot_sipp_2023}, it was shown that this thickness corresponds to the displacement thickness at the attachment-line. For the present configuration, we have $\Delta=9.71 \times 10^{-5}$. % We will also introduce below the curvilinear abscissa $\chi$ along a streamline of the external base-flow velocity field, just outside of the boundary layer, while $ b $ is normal to the plane $(\chi,\eta)$.

\begin{comment}
This value of the amplification ($ln(\sigma)\approx9$) is typical of the amplifications obtained in local REF ? analysis. 
\end{comment}

\subsection{Wall-displacement resolvent modes}

%\subsection{Study of the optimal roughness and response}

\begin{figure}
\centering
\includegraphics[trim =02mm 00mm 0mm 0mm, clip,width=0.7\textwidth]{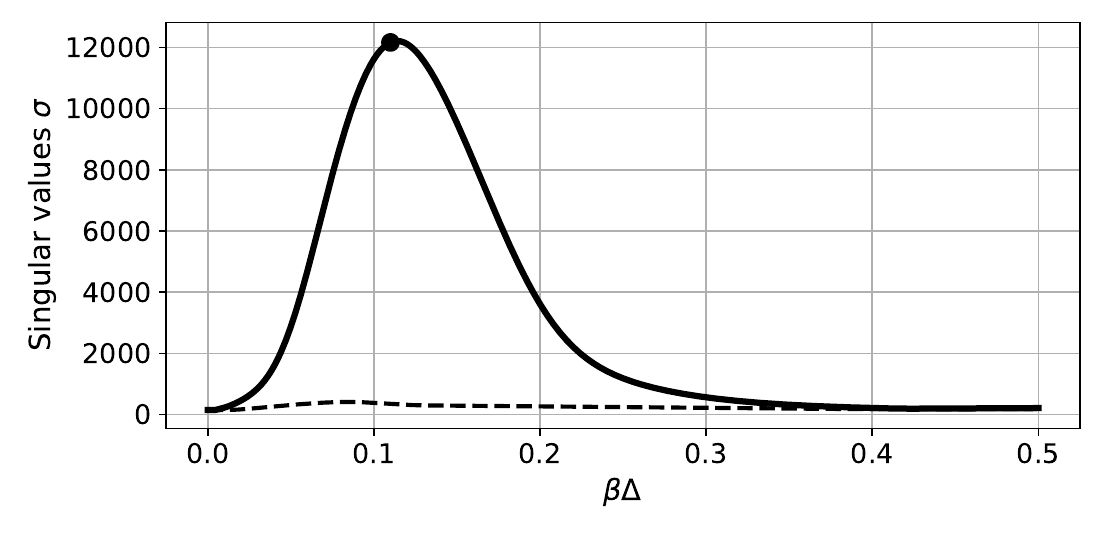}
\vspace{-0.5cm}
\caption{First two singular values: $\sigma_1$ in solid line and $\sigma_2$ in dashed line. The largest singular value ($\beta\Delta=0.11$) is marked with a circle.}
\label{fig:SV}
\end{figure}

Figure \ref{fig:SV} shows the first two singular values $\sigma_1$ and $\sigma_2$
as a function of the spanwise wavenumber $0 \leq \beta\Delta \leq 0.5 $. For values of $\beta\Delta$ between $0.05$ and $0.2$, the first singular value is significantly higher than the second one, making the resolvent operator be close to rank 1. The second singular value has a maximum of $\sigma_2=415$ reached for $\beta\Delta=0.08$, while the maximum of the first singular value is achieved for $\beta\Delta=0.11$ where $\sigma_1=12162$. The singular value $\sigma_1$ for $\beta\Delta=0.11$, highlighted by the black circle, corresponds to the configuration that is studied in more detail in the following. \\
%%%%%%%%%%%%%%%%%%%%%%%%%%%%%%%%%%%%%%%%%%%%%%%%%%%%
 %%%%%%%%%%%%%%%%%%%%%%%%%%%%%%%%%%%%%%%%%%%%
\subsubsection{Optimal roughness and response for $\beta\Delta=0.11$}\label{part:kz0.11}

\begin{figure}[h!]
\begin{center}
\begin{tabular}{rlrl}
(a) & & (b) & \\
& \hspace{-0.5cm}\includegraphics[trim =02mm 42mm 18mm 35mm, clip,width=0.45\textwidth]{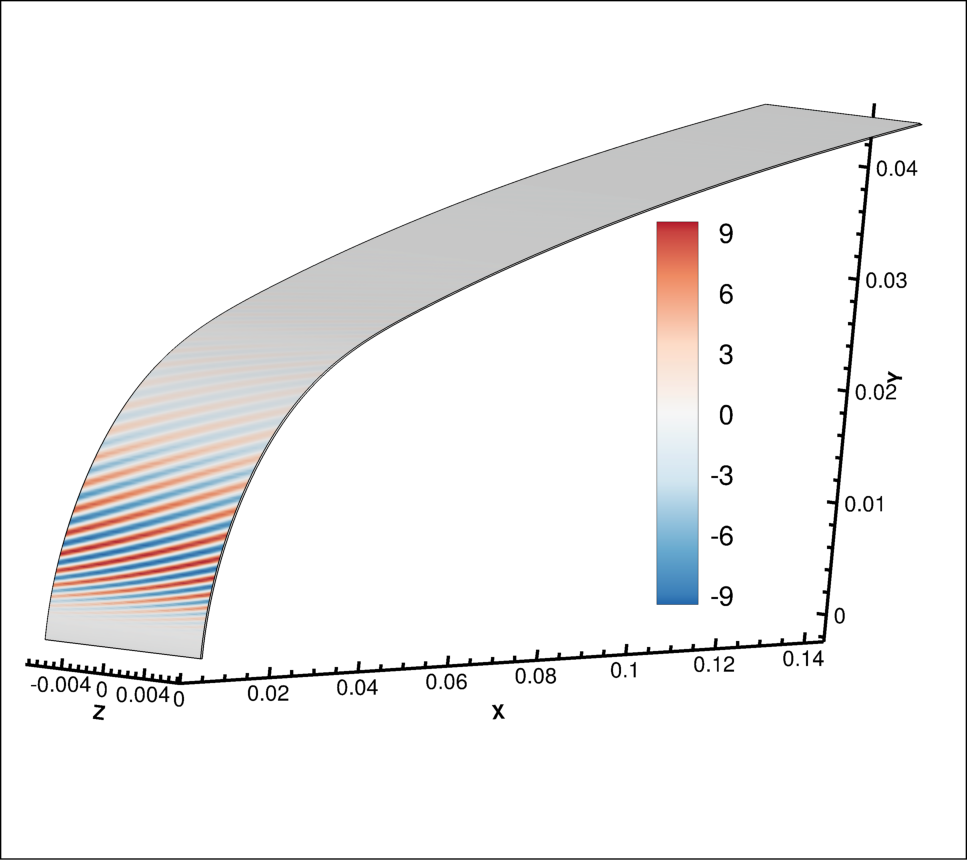} & & \hspace{-0.5cm}\includegraphics[trim =26mm 72mm 26mm 55mm, clip,width=0.45\textwidth]{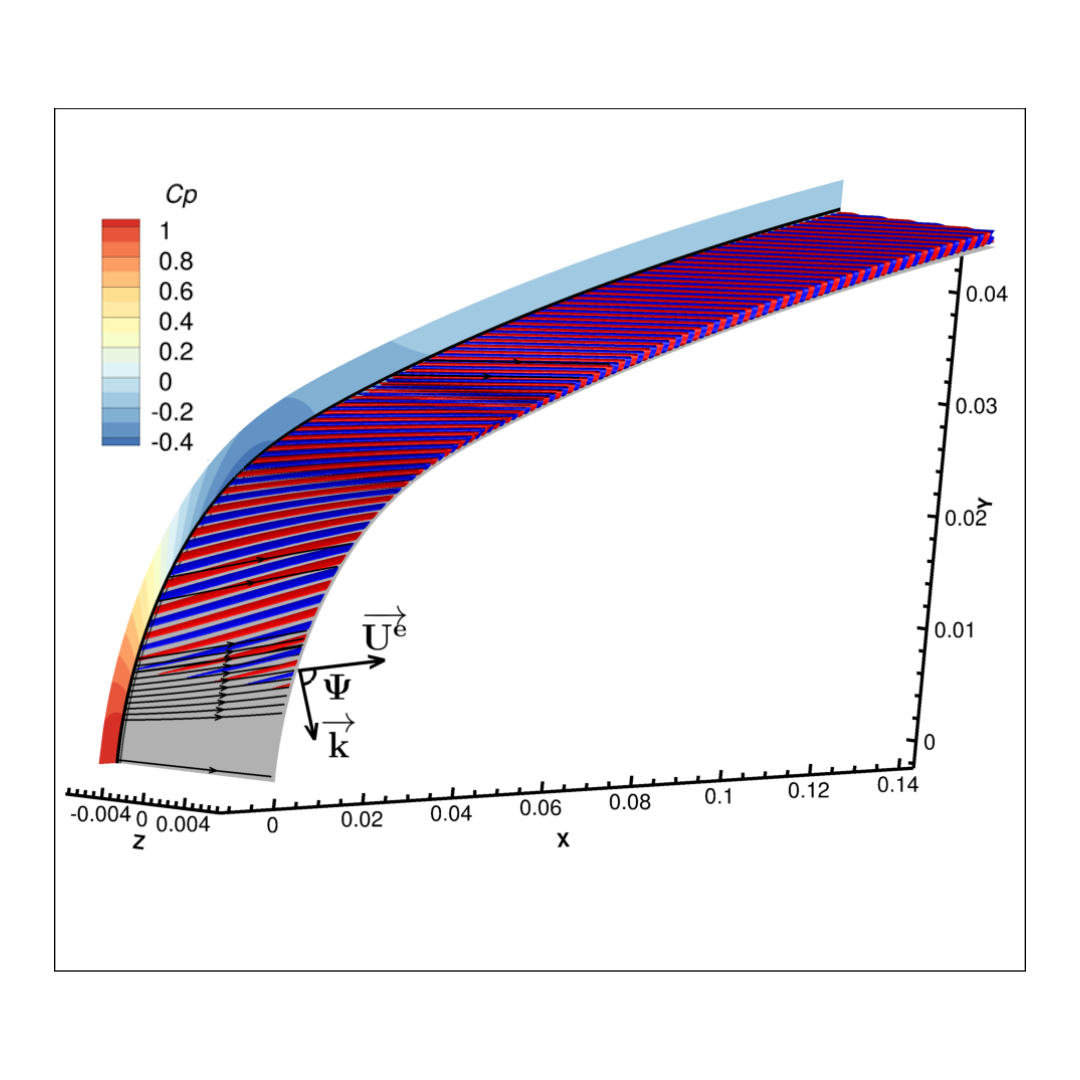}
\end{tabular}
\end{center}
 \vspace{-0.2cm}
\caption{Spatial structure of the real part of: the optimal roughness $\Re(\hat{h}_{0.11/\Delta,1}(s)e^{i\beta z})$ $(a)$ and the $z$-velocity of the response $\Re\left(\hat{u}_{0.11/\Delta,1,z}(x,y)e^{i\beta z}\right)$$(b)$. 2 iso-surfaces at $\pm 0.1$ times the absolute maximum are represented in red and blue. Pressure coefficient $C_p$, boundary layer thickness $\delta_{99}$ (black line), and potential streamlines (black arrow lines) are shown. An example of wavevector and $\Psi$ angle is also displayed.}
\label{fig:Response3D}
\end{figure}

We now analyse the spatial structure of the optimal response and roughness for the spanwise wavenumber close to the largest singular value, i.e. $\beta\Delta=0.11$. In figure \ref{fig:Response3D} are represented the iso-surfaces of the real part of the optimal roughness and of the spanwise velocity of the optimal response. 
The optimal roughness is located at the beginning of the leading edge and is oriented in a direction close to the external streamline. The optimal response has a large magnitude on the whole leading edge except at the attachment-line. It consists in steady vortices whose axes are nearly parallel to the external streamlines.

To help discern the type of instability, as commonly done in local stability approaches \citep{Arnal2000173}, we introduce the $\Psi$ angle between the local planar wave vector $\vec{k}(s)=[{k}_s(s),k_z]$ %($k_\chi(s)=\|\vec{k}(s)\|$)
of the mode and the local direction of the external baseflow streamline: $\Psi=\textrm{angle}(\vec{k}(s),\vec{U}^e(s))$.
Such a planar wave-vector may be approximated as follows: if $\hat{u}(x,y)e^{i\beta z}$ is a component of the perturbation, then $ (k_s,k_z)=(\partial_{s}  \phi,\beta) $ where $\phi(s,\eta)=\arg{\hat{u}(s,\eta)}$.
The choice of the component and wall normal distance $ \eta $ does not matter as long as the flow is weakly non-parallel (condition for the existence of such a local wave-vector).
Here we used the $ \hat{u}_y$-component and $ \eta=\delta_{99}/2$ where $\delta_{99}$ is the wall normal distance such as $U_\chi(\delta_{99})=0.99U_\chi^e$.
The same technique may be used to obtain a $ \Psi $ angle for the optimal roughness $ \hat{h}(s)e^{i\beta z} $.

\begin{figure}[h!]
\begin{center}
\begin{tabular}{rlrl}
(a) & & (b) & \\
& \hspace{-0.9cm}\includegraphics[trim =10mm 0mm 10mm 0mm, clip,width=0.49\textwidth]{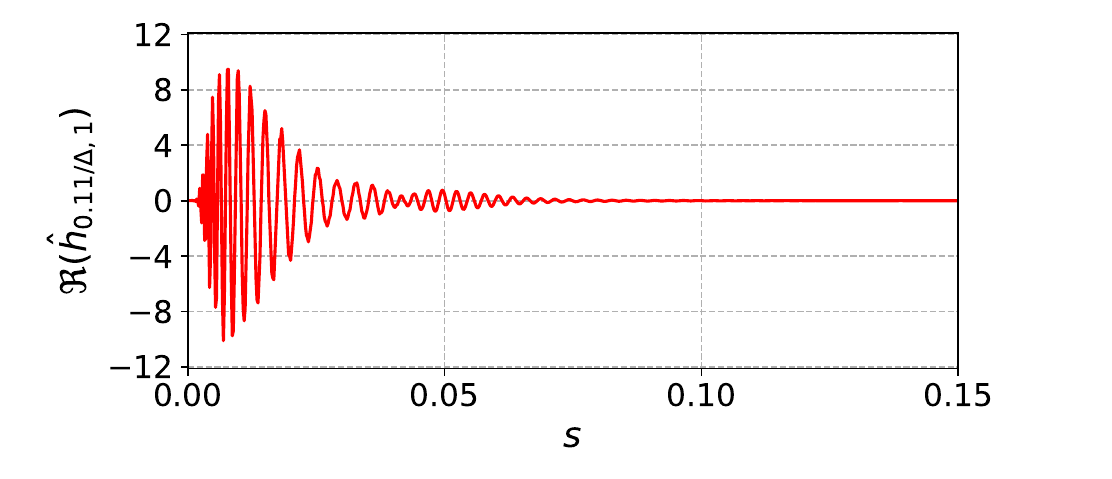} & & \hspace{-0.9cm}\includegraphics[trim =-0mm 0mm 0mm -0mm, clip,width=0.49\textwidth,height=0.235\textwidth]{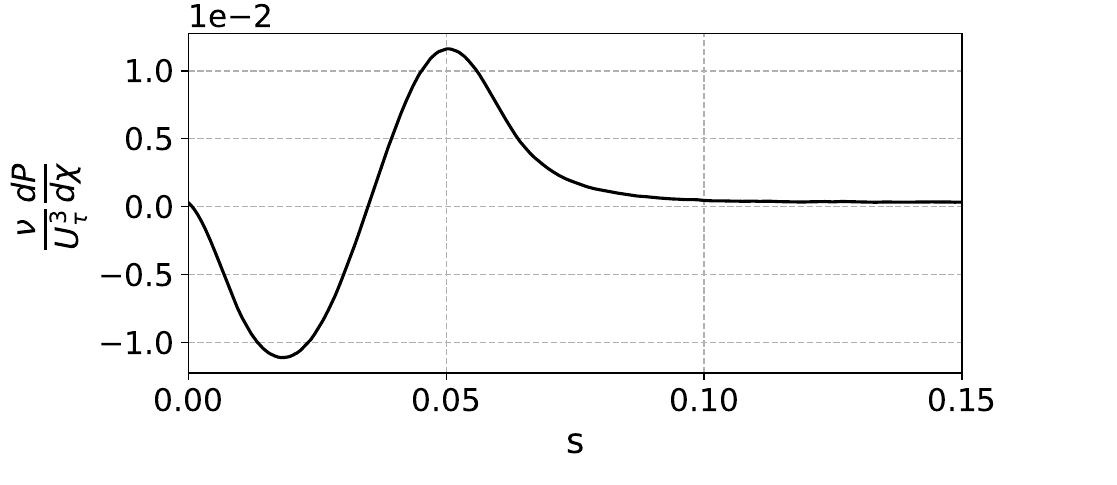} \\
(c) & & (d) & \\
& \hspace{-0.9cm}\includegraphics[trim =10mm 0mm 10mm 0mm, clip,width=0.49\textwidth]{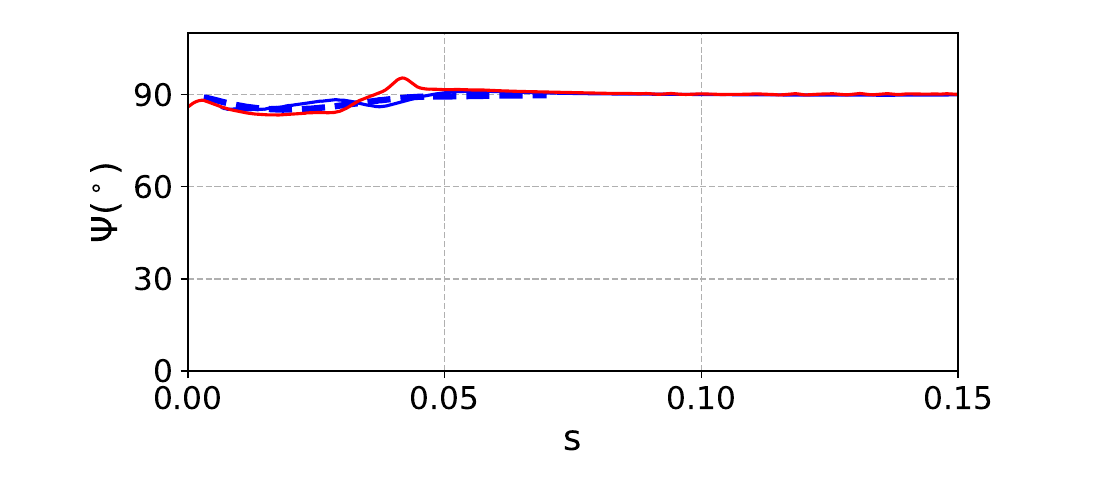} & & \hspace{-0.9cm}\includegraphics[trim =0mm 0mm 0mm 0mm, clip,width=0.49\textwidth,height=0.235\textwidth]{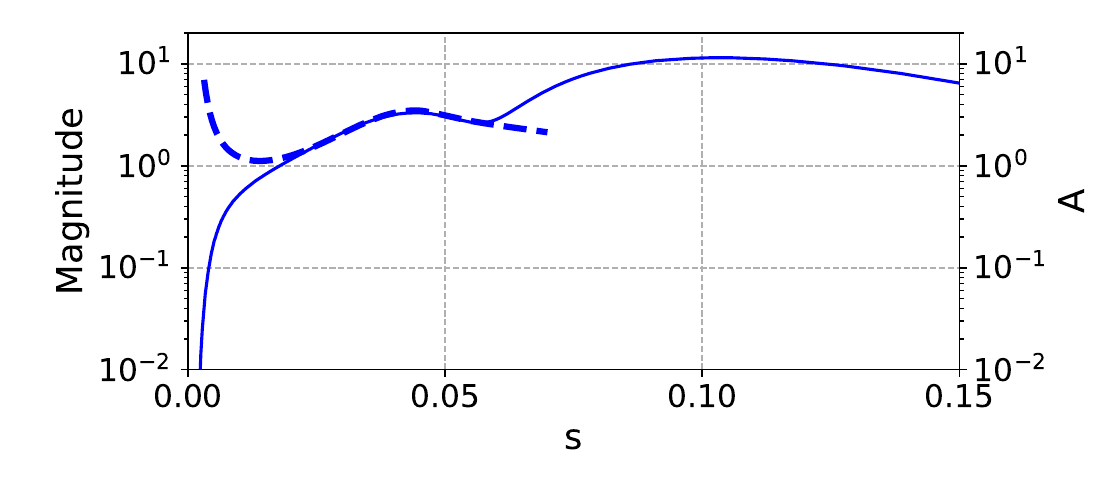}
\end{tabular}
\end{center}
 \vspace{-0.2cm}
\caption{Curvilinear evolution of $(a)$: the optimal roughness. $(b)$: the streamwise pressure gradient made non-dimensional with friction velocity $U_\tau = (\nu \partial_\eta U_{\chi}(\eta=0))^{0.5}$ and kinematic viscosity $ \nu $. $(c)$: the $\Psi$ angle of the optimal roughness (red) and response (blue). $(d)$: the magnitude of the optimal response. The optimal perturbation obtained with the global resolvent (in solid line) and the mode calculated by a local stability analysis (in dashed line) are represented.}
\label{fig:ForcingResponseKz0.11}
\end{figure}

 The magnitude and orientation of the optimal roughness as well as of the associated perturbation are displayed in figure \ref{fig:ForcingResponseKz0.11} and the pressure gradient.
 
The real part of the optimal roughness is plotted in figure \ref{fig:ForcingResponseKz0.11}$(a)$. Its amplitude is close to zero at both extremities of the domain and reaches its maximum magnitude at $s\approx0.008$. It also has a second (weaker) local maximum at $s=0.05$ with a second amplification region starting at $s=0.04$. 
The curvilinear evolution of the $\Psi$ angle of the optimal roughness is represented in red in figure \ref{fig:ForcingResponseKz0.11}$(c)$. We observe that it remains close to $90^\circ$ on the whole domain, with small variations up to $s=0.04$.
Concerning the optimal perturbation, the magnitude of the mode as a function of $s$ is plotted in figure \ref{fig:ForcingResponseKz0.11}$(d)$. The magnitude $d_{\boldsymbol{\hat{u}}}(s)$ is defined as
$ d_{\boldsymbol{\hat{u}}}(s)=\sqrt{\int_0^{L_\eta} \| \boldsymbol{\hat{u}}(s,\eta) \|^2 d\eta}$, where $ L_\eta=45\Delta$.
We notice a weak magnitude at $s=0$, a strong amplification from $s=0$ to $s=0.01$, a decrease around $s=0.05$ and finally a second amplification from $s=0.057$ up to $s=0.1$.
The evolution according to $s$ of its $\Psi$ angle is displayed in blue in figure \ref{fig:ForcingResponseKz0.11}$(c)$ and we also notice a value close to $90^\circ$ on the whole domain. Figure \ref{fig:ForcingResponseKz0.11}$(b)$ represents, as a function of $s$, the pressure gradient scaled using $U_\tau = (\nu \partial_\eta U_{\chi}(\eta=0))^{0.5}$ and $ \nu $. In the ONERA-D case, the streamwise pressure gradient is negative up to $s=0.035$, then positive until the limit of the domain, with a flattening around $s=0.09$. This pressure-gradient changeover is typical of a flow on a swept wing and leads to the existence of two inflection points in the crossflow velocity profile for some values of $s$ \citep{wassermann2005transition,Arnal2000173}. A negative pressure gradient is favourable to the development of crossflow waves, which accounts for the increase in the magnitude of the response at the beginning of the domain, while a positive pressure gradient is generally responsible for the growth of Tollmien-Schlichting waves. Additional results about the baseflow for a configuration close the current one are presented in \cite{kitzinger_leclercq_marquet_piot_sipp_2023}. 

Based on the characteristics of the optimal response, namely a high magnitude at the leading edge but not in the close vicinity of the attachment-line and a $\Psi$ angle close to $90^\circ$, we can conclude that the optimal response is a cross-flow type mode. This is consistent with previous observations of the modes appearing in the case of swept wings with wall roughness \citep{Saric2003}.
Moreover, the fact that, having an identical spanwise wavenumber, the optimal roughness and the associated perturbation have an almost equal curvilinear evolution of the $\Psi$ angle shows that they also have a very close evolution of their curvilinear wavenumber $k_s(s)$, as observed by \citet{tempelmann2012swept1}.

A direct link between the double amplification of the optimal roughness and the associated response was not identified. Indeed, when calculating the response associated with a roughness height equal to the optimal roughness height for $s<0.04$ and to $0$ beyond, the same second amplification was observed. This is also shown by the roughness studied in section \ref{results:squareWave}.
 
\begin{figure}
\centering
\includegraphics[trim =02mm 00mm 0mm 0mm, clip,width=0.5\textwidth]{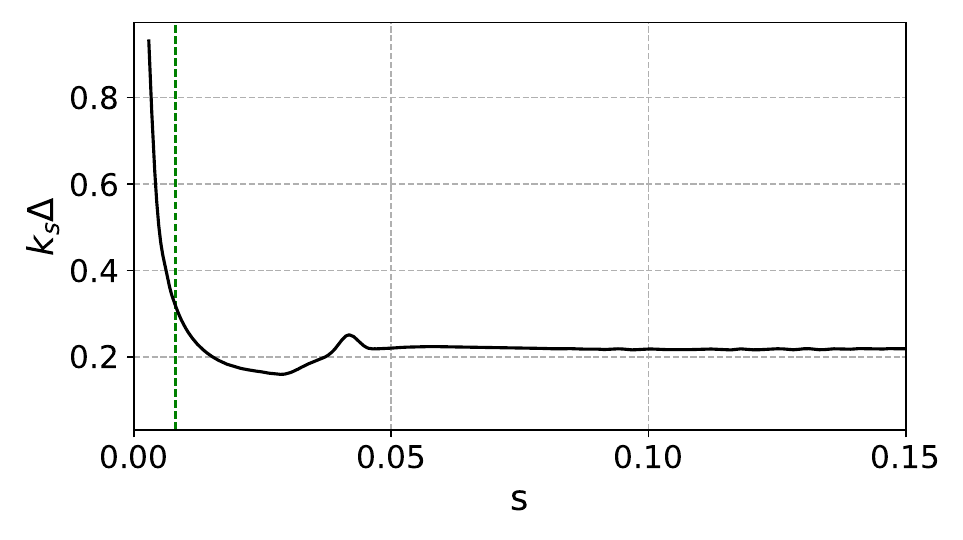}
\caption{Curvilinear evolution of $k_s\Delta$ of the optimal roughness with $ \beta \Delta=0.11$. The location of the maximum magnitude of the optimal roughness is shown (green vertical dashed line).}
\label{fig:ksDelta_ForcingOpt_kz0.11}
\end{figure}

%It should also be noted that the response amplification observed at the very beginning of the domain in figure \ref{fig:ForcingResponseKz0.11}$(d)$ does not correspond to an amplification as conceived in a local stability analysis.
We now compare the magnitude and the $\Psi$ angle of the optimal response with the results obtained using a local stability analysis (figure \ref{fig:ForcingResponseKz0.11}$(c,d)$). 
The local stability analysis considers eigenmodes sought in the $(s,\eta,z)$ reference frame under the form $q=\hat{q}(\eta)e^{i(\alpha s + \beta z - \omega t)}$. The spatial stability analysis in the $s$-direction is solved for fixed $\beta$ and $\omega$ real values. The local stability code solves the one-dimensional differential eigenvalue problem with a high-order scheme. The parallel flow assumption is used, and the flow computed by the boundary-layer solver is used as the baseflow, to avoid interpolation errors from the FEM mesh. In the local stability analysis framework, the $\Psi$ angle is directly derived from the real parts of $\alpha$ and $\beta$ and the knowledge of the inviscid streamwise direction at each chordwise location. The spatial amplification $A(s)$ is represented in figure \ref{fig:ForcingResponseKz0.11}$(d)$ and is defined as $\ln(A(s)/A_0)=\int_{s_0}^s-\Im{(\alpha (s))}ds$ \citep[see for instance][for reviews on local stability approach]{Arnal2000173,reed1996linear} with the initialisation at $s_0=0.003$ and $A_0$ arbitrarily chosen such as $A(s)$ fits the magnitude of the optimal response. We observe that the $\Psi$ angle and the amplification for $0.018<s<0.056$ match very well.
For $s<0.013$, the mismatch is due to the fact that the optimal response is triggered by a roughness, which is not taken into account in the spatial stability analysis. % After this location, the optimal roughness is weak, which then allows a good agreement between the two stability approaches.
The position of the beginning of the growth phase is the position of branch I and coincides closely with the maximum magnitude of the associated roughness ($s=0.008$). This is in good agreement with the literature \citep{tempelmann2012swept1,sipp2013characterization,choudhari1994roughness}.
Finally, the second amplification of the mode from the resolvent analysis ($s>0.057$) could not be captured by the local stability analysis. This may be due to several reasons. First, the second amplification may be caused by a non-modal spatial growth, which is not captured by examining only the most unstable mode of the local stability analysis. Secondly, the assumptions of flow parallelism and no surface curvature used in the local stability analysis are also limiting and may account for this deviation. It is not obvious whether a PSE method, which takes into account weak non-parallelism but does not capture non-modal mechanisms \citep{towne2019critical}, would be able to recover this second amplification.
%\textcolor{green}{These issues show that the optimal perturbation could not have been calculated so easily using a PSE.} \textcolor{red}{Furthermore, this perturbation could not have been calculated either with the method used by \cite{tempelmann2012swept1,thomas2017predicting} because the latter requires the perturbation to correspond to an identical local stability mode at the inflow and outflow of the domain.}

% \begin{figure}[h!]
% \begin{center}
% \begin{tabular}{rlrl}
% (a) & & (b) & \\
% & \hspace{-1cm}\includegraphics[trim =0mm 0mm 0mm 00mm, clip,width=0.5\textwidth]{CompLocal_Magnitude.pdf} & & \hspace{-1.2cm}\includegraphics[trim =-0mm 0mm 0mm 0mm, clip,width=0.5\textwidth]{CompLocal_Psi.pdf}}
% \end{tabular}
% \end{center}
%  \vspace{-0.2cm}
% \caption{Comparison of the curvilinear evolution of the $\Psi$ angle $(b)$ and of the magnitude and spatial amplification $(a)$ of the optimal perturbation obtained with the global resolvent (in solid line) and the mode calculated by a local stability analysis (in dashed line).}
% \label{fig:CompLocGlob}
% \end{figure}

The evolution of the curvilinear wavenumber $k_s$ with respect to $s$ is represented in figure \ref{fig:ksDelta_ForcingOpt_kz0.11} and the position of the maximum magnitude of the optimal roughness is indicated with a green vertical dashed line. The wavenumber $k_s$ decreases from the attachment-line till $s=0.03$ where $k_s\Delta=0.16$. At the location of maximum magnitude $s=0.008$, we get $k_s\Delta=0.32$. Afterwards, it reaches a local maximum of $k_s\Delta=0.25$ where the second amplification region begins and is then about $0.22$ up to the end of the domain. Moreover, since the spanwise wavenumber is fixed and $\Psi\approx90^\circ$ on the whole domain, the curvilinear evolution of $k_s$ is closely related to the orientation of the external streamlines.

\subsubsection{Optimal roughness and perturbations for $0.05 \le \beta\Delta \le 0.3$}

The optimal roughness for $\beta\Delta\in[0.05,0.3]$ have qualitatively similar spatial structures as the one at $\beta\Delta=0.11$ with two magnitude maxima and zero values at the extremities of the domain as well as a curvilinear wavenumber which varies with $s$. 

\begin{figure}[h!]
\begin{center}
\begin{tabular}{rlrl}
(a) & & (b) & \\
& \hspace{-0.5cm}\includegraphics[trim =0mm 0mm 0mm 00mm, clip,width=0.45\textwidth]{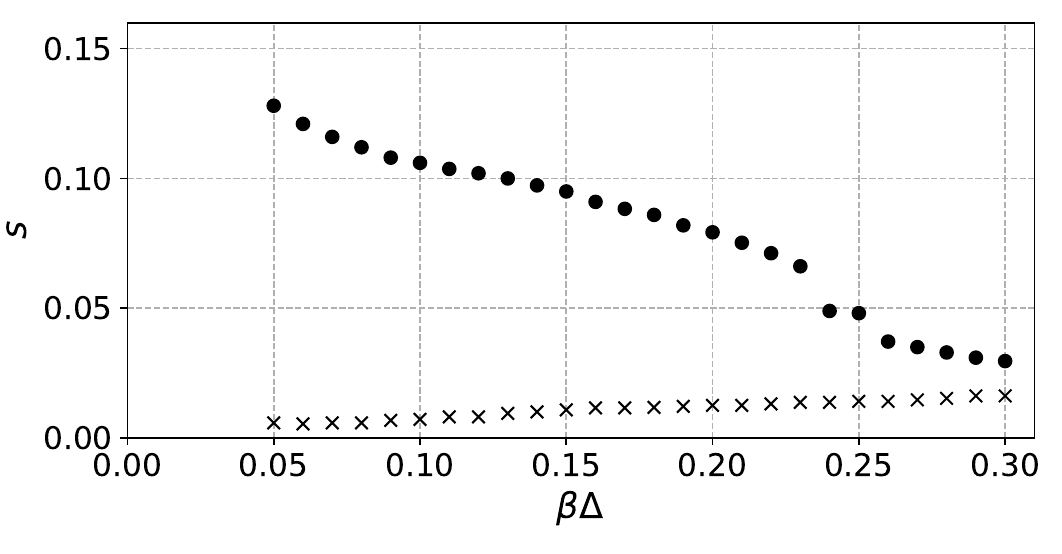} & & \hspace{-0.5cm}\includegraphics[trim =0mm 0mm 0mm 00mm, clip,width=0.45\textwidth]{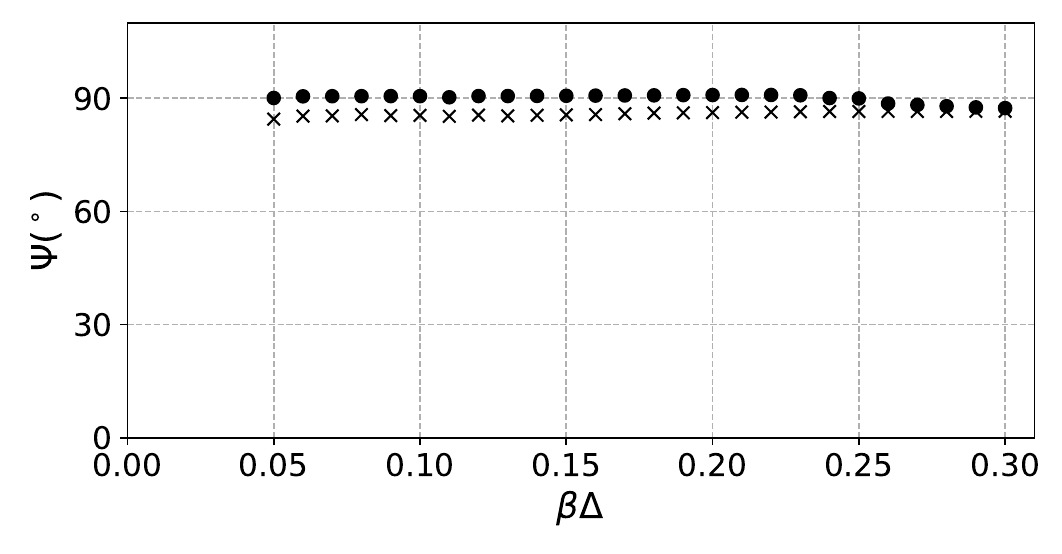} 
\end{tabular}
\end{center}
 \vspace{-0.2cm}
\caption{$(a)$: Curvilinear position of the maximum magnitude of the optimal roughness and response as a function of $ \beta \Delta$. $(b)$: Local $\Psi$ angle of the optimal roughness and response at the location of the maximum magnitude. The values related to the optimal roughness and responses are respectively depicted by crosses and circles.}
\label{fig:s_Psi_kz}
\end{figure}

In figure \ref{fig:s_Psi_kz}$(a)$ are plotted according to $\beta\Delta$ the positions of the maximum magnitude of the optimal roughness and perturbation for the different spanwise wavenumbers. Concerning the optimal roughness, the position of the maximum tends to get further away when increasing $\beta\Delta$, moving from $s\approx0.006$ to $s\approx0.016$. For all spanwise wavenumber values, the optimal roughness is located relatively close to the attachment-line, thus corroborating the conclusions of previous studies \citep{Radeztsky1999,thomas2017predicting}. Contrary to the optimal roughness, the position of the maximum magnitude of the optimal response tends to get closer to the attachment-line with the increase of the spanwise wavenumber, moving from $s=0.13$ for $\beta\Delta=0.05$ to $s=0.022$ for $\beta\Delta=0.30$.

In figure \ref{fig:s_Psi_kz}$(b)$ are represented the $\Psi$ angles of the optimal roughness and response at the curvilinear location of the maximum magnitude as a function of $\beta\Delta$. We note that the $\Psi$ angle remains almost constant with the evolution of the spanwise wavenumber with values close to $90^\circ$.

These observations on the position of the maximum of the magnitude of the optimal response as well as on the value of the associated $\Psi$ angle tend to show that the optimal response is of cross-flow type for $\beta\Delta\in[0.05,0.3]$.

\begin{comment}
Elargir à l'étude des réponses avec des commentaires dessus : We also observe that the responses have a similar structure characteristic of CF.

In figure \ref{fig:ks_kz}$(b)$ is represented the curvilinear wavenumber $k_s\Delta$ as a function of $\beta\Delta$. We observe an almost linear evolution with a coefficient ?? of this wavenumber as a function of the wavenumber in the $z$-direction. (A VERFIER AVEC UUNE REGRESSION LINEAIRE A COMPARER AVEC UNE REGRESSION QUADRATIQUE)
\end{comment}

\subsection{Flow response to specific roughness}\label{part:ResponseGivenForcing}

In the present section, we assess the performance of the formulation \eqref{eq:ProjBasis} associated with the approximation \eqref{eq:LowRankProj} to reconstruct the response triggered by localised and non-localised roughness.

\subsubsection{Case of square wave roughness localised in the curvilinear $s$-direction and harmonic in the spanwise $ z$-direction}\label{results:squareWave}

\begin{figure}
\begin{center}
\begin{tabular}{rlrl}
(a) & & (b) & \\
& \hspace{-0.5cm}\includegraphics[trim =04mm 65mm 32mm 40mm, clip,width=0.45\textwidth]{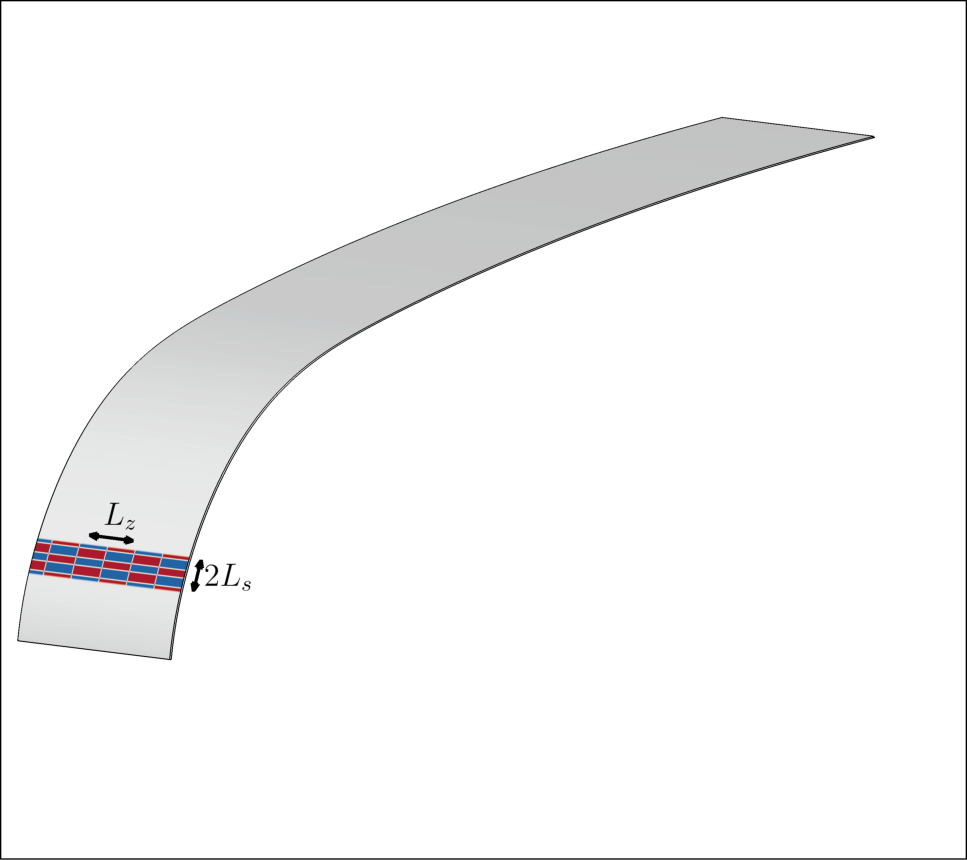} & & \hspace{-0.5cm}\includegraphics[trim =04mm 40mm 10mm 45mm, clip,width=0.45\textwidth]{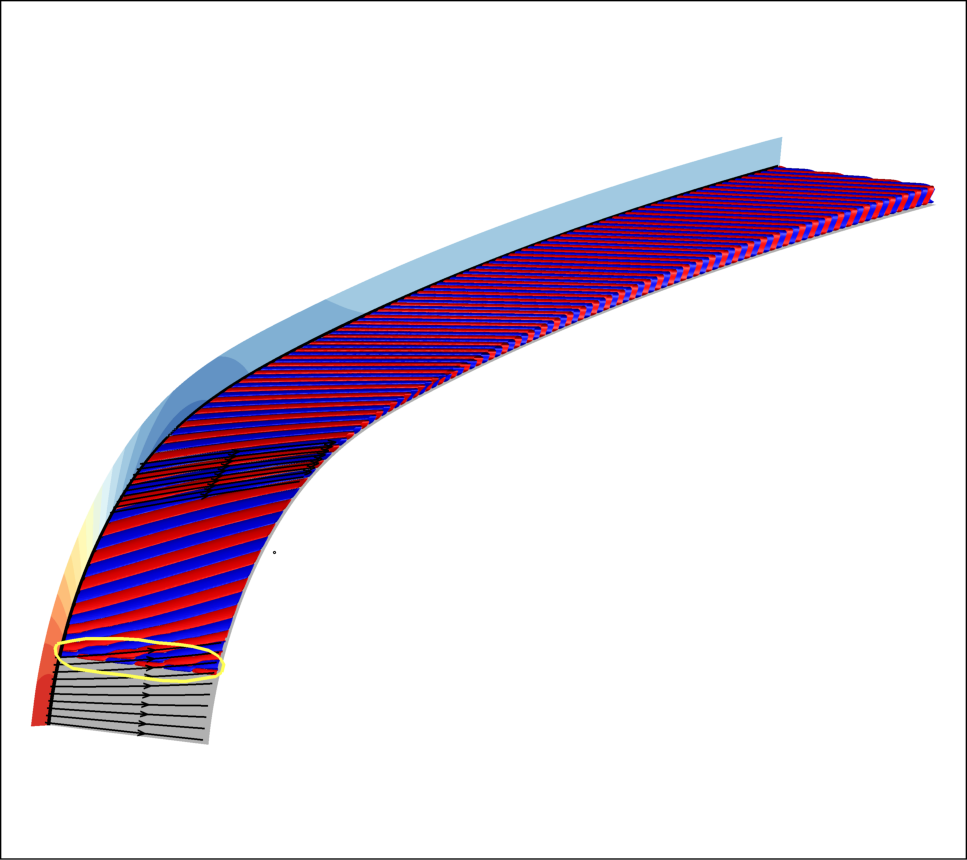}
\end{tabular}
\end{center}
\caption{$(a)$: Shape $ h(s,z)/H$ of the square wave roughness with $L_z/\Delta=2\pi/0.11\approx 57.1$ and $L_s/\Delta=2\pi/0.32\approx 19.6$ and localised at $s_0=0.008$. The positive and negative values are represented in red and blue, respectively. $(b)$: Associated response calculated solving the LNSE with the same representation as \ref{fig:Response3D}. The yellow ellipse indicates the area where a difference with the response calculated using the approximation \eqref{eq:projCreneauApprox} occurs.}
\label{fig:creneau3D_bouge}
\end{figure}

We consider a roughness periodic (characteristic size $ L_z=2\pi/\beta_0$) in $z$ and which is compact (characteristic size $ L_s$) and localised in the curvilinear direction (around $ s_0 $):
\begin{equation} \label{eq:separate}
    \frac{h(s,z)}{H}=\frac{1}{2}{h}_s\left(\frac{s-s_0}{L_s}\right){h}_z\left(\frac{z}{L_z}\right),
\end{equation}
where
\begin{eqnarray}
{h}_{s}&=&-\mathcal{H}_{-1}+2\mathcal{H}_{-3/4}-2\mathcal{H}_{-1/4}+2\mathcal{H}_{1/4}-2\mathcal{H}_{3/4}+\mathcal{H}_{1} \\
{h}_z&=&-1+2\mathcal{H}_{0} \mbox{ for } z\in [-1/2;  1/2[ \;\;\;\;\mbox{and} \;\;\;\; {h}_z(z+1)={h}_z(z) \;\; \forall z,
\end{eqnarray}
function $ \mathcal{H}_{\xi}$ referring to the Heaviside function with the discontinuity at $ \xi$. 
The roughness exhibits a square wave shape of peak-valley distance $H$, zero mean, both in the chordwise direction $s$ and the spanwise direction $z$. Its spanwise wavenumber $\beta_0$ is chosen so as to coincide with the most amplified spanwise wavenumber optimal roughness ($\beta_0\Delta=0.11$). We picked $L_s$ close to the local wavelength $ L_s=2\pi/k_s=2\pi \Delta /(k_s\Delta=0.32)$ of this optimal roughness at its maximum magnitude occurring at $s_0=0.008$ (see figure \ref{fig:ksDelta_ForcingOpt_kz0.11}). %The height $ H $ is chosen small enough with respect to the boundary layer size: $H/\Delta=0.01$.
The roughness centred at $ s_0=0.008 \approx 82 \Delta $ is represented in figure \ref{fig:creneau3D_bouge}. In the following, we will also consider this same roughness but moved around $ s_0=0.016\approx 165 \Delta$ and $ s_0=0.056 \approx 577 \Delta$: since we keep the same $ L_s $, these roughness are not anymore optimal with respect to the local optimal roughness wavenumber, which is respectively $ k_s\Delta=0.19$ and $ k_s\Delta=0.22$ at these new locations.   

Equation \eqref{eqn:Fourier} for the wall roughness becomes $ {h}(s,z)=(H/2){h}_s((s-s_0)/L_s)\sum_{m=-\infty}^{+\infty} \hat{h}_{z,m}e^{im\beta_0 z}$,
where $ \hat{h}_{z,m}$ is the Fourier transform of a zero-mean square wave, so that $\hat{h}_{z,0}=0$ and $ |\hat{h}_{z,m}| \sim m^{-1} $. 
\begin{comment}
Since the chosen forcing has spatial characteristics close to an optimal forcing associated with a dominant singular value, we neglect the contributions of the following singular values. and only keep 4 harmonics in $ m $. Thus, by considering only the four first harmonics, equation \eqref{eq:projCreneauReponse} becomes:

\begin{equation}
\begin{split}
     h(s,z)&=H\sum\limits_{m=-\infty}^{+\infty}c_{z,m}\hat{h}_s(s)e^{im\beta_0 z}\stackrel{\hat{h}_z\in\mathbb{R}}=Hc_{z,0}\hat{h}_s(s)+2H\sum\limits_{m=1}^{+\infty}\Re(c_{z,m}\hat{h}_s(s)e^{im\beta_0 z})
\end{split}
\end{equation}
with $c_{z,m}$ the spanwise Fourier coefficient of the $m^{th}$ harmonic.

\begin{equation}
  \begin{split}\label{eq:projCreneauApprox}
    \boldsymbol{u}(x,y,z)&\approx 2H\sum\limits_{m=1}^{4}\Re\left(c_{z,m}\sigma_{1,m}\langle\hat{h}_s(s),\hat{h}_{1,m}(s)\rangle_w\boldsymbol{\hat{u}}_{1,m}(x,y)e^{im\beta_0 z}\right)
  \end{split}
  \end{equation}.
\end{comment}
 Considering an approximation with the dominant singular value and 4 harmonics, equation \eqref{eq:ProjBasis} for the response provides the following explicit form of the flow response:
\begin{equation}   \label{eq:projCreneauApprox} \frac{\boldsymbol{u}(x,y,z)}{H}\approx
\sum_{m=1}^{4}\Re\left(\hat{h}_{z,m}\sigma_{1,m}\left\langle\hat{{h}}_{1,m}(s),{h}_s\left(\frac{s-s_0}{L_s}\right)\right\rangle_w\boldsymbol{\hat{u}}_{1,m}(x,y)  e^{im\beta_0 z}\right).
  \end{equation} 
  The local mean fluctuation rate $ \langle\| \boldsymbol{u} \|\rangle_z(s,\eta)/H$ of the perturbation may then be obtained by an average of the kinetic energy in the spanwise direction, see equation \eqref{eq:mean2}. 

\begin{figure}[h!]
\begin{center}
\begin{tabular}{rl}
(a) &\\
& \hspace{-0.5cm}\includegraphics[trim =-0mm 0mm 0mm 0mm, clip,width=0.95\textwidth]{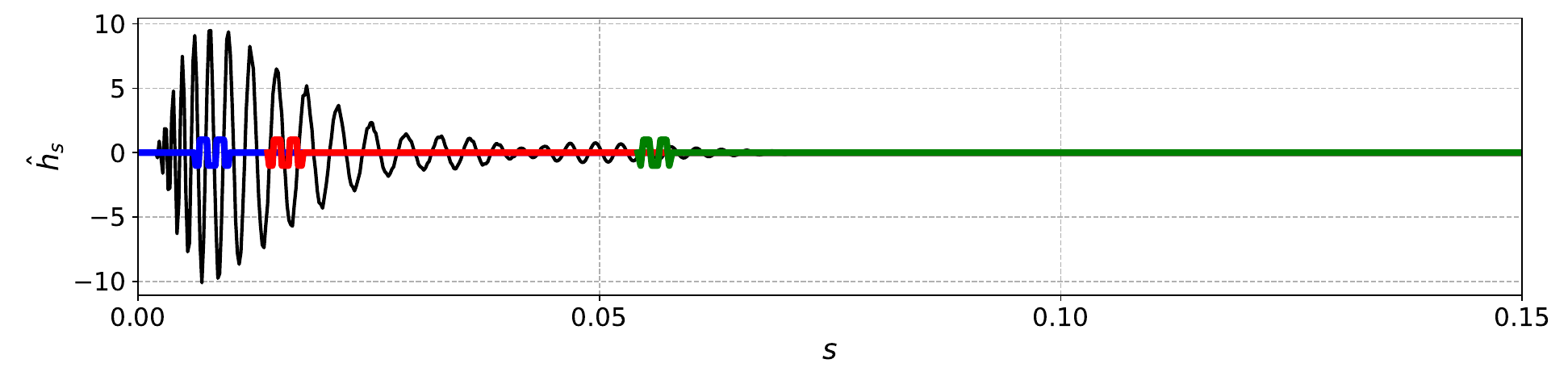} \\
(b) & \\
& \hspace{-0.5cm}\includegraphics[trim =0mm 0mm 0mm 0mm, clip,width=0.95\textwidth]{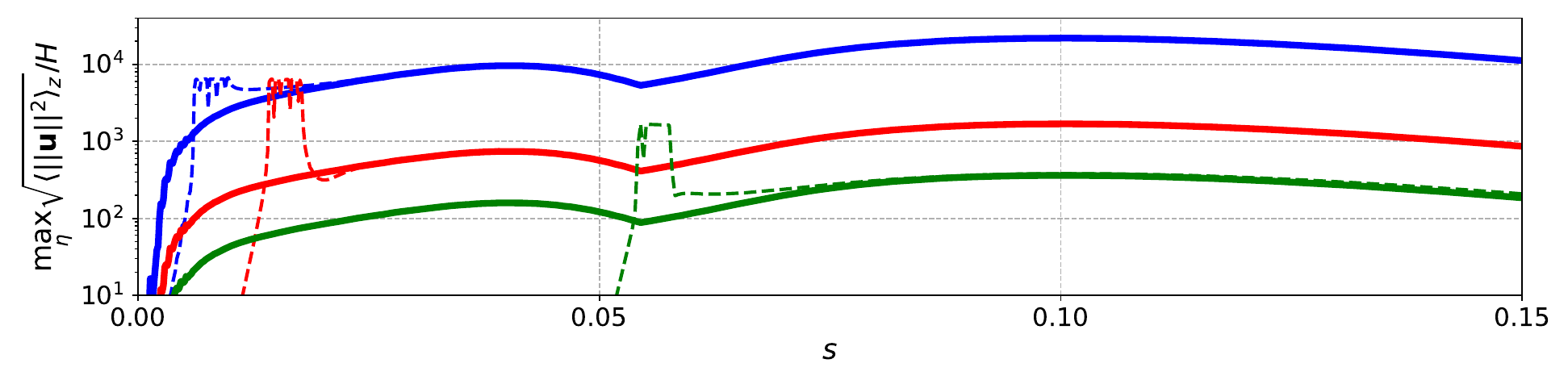}
\end{tabular}
\end{center}
 \vspace{-0.2cm}
\caption{$(a)$: Representation of the optimal roughness (black) and the shape of the square wave roughness $ h_s((s-s_0)/L_s) $ localised at $s_0=0.008$ (blue), $s_0=0.016$ (red) and $s_0=0.056$ (green) and $L_s/\Delta \approx 18.5$. The square wave roughness in $(a)$ have been magnified  by a factor $2$ for visualisation purposes. $(b)$: Curvilinear evolution of the maximum mean fluctuation rate $ \max_\eta \sqrt{\langle\| \boldsymbol{u} \|^2\rangle_z}/H$ of the responses by resolution of the equations \eqref{ResolventDiscret} (dashed lines) and by using the approximation in \eqref{eq:projCreneauApprox} (solid lines).}
\label{fig:CreneauPartiel}
\end{figure}

We have considered three locations of the roughness in figure \ref{fig:CreneauPartiel}: $s_0=0.008$ (blue), $s_0=0.016$ (red) and $s_0=0.056$ (green).
They are represented in figure $(a)$ along with the optimal roughness (black). In figure $(b)$ are represented the curvilinear evolution of the maximum mean fluctuation rate $\max\limits_\eta \sqrt{\langle\| \boldsymbol{u} \|^2\rangle_z}/H$ of the response obtained by solving the exact equation \eqref{ResolventDiscret} (dashed lines) and by using the approximation formula defined in \eqref{eq:projCreneauApprox} (solid lines). 
A good agreement between the responses is found downstream of the roughness. Hence, if the roughness is sufficiently small not to trigger transition in its vicinity (natural transition% and not by-pass transition
), the downstream evolution of the fluctuation rate is well captured by few singular values/harmonics. 
The magnitude peak at the location of the roughness is more noticeable when the response has low energy, so that the maximum magnitude of the most critical roughness is still well captured by the low-rank approximation \eqref{eq:projCreneauApprox}.

%\textcolor{green}{In addition, we show that the resolution of the LNSE (or the consideration of a sufficient number of sub-optimal responses) makes it possible to calculate the spatial structure of the response at the location of the roughness and upstream of it, which is not possible using the PSE }\textcolor{red}{ or the adjoint based method used in \cite{tempelmann2012swept1,thomas2017predicting}. } 

Depending on the position of the roughness, the term $\langle\hat{h}_{1,m}(s), {h}_s((s-s_0)/L_s)\rangle_w$ is modified. The latter is higher when the roughness is located where the optimal roughness has a strong magnitude and a similar curvilinear periodicity. That's why the response computed with equation \eqref{eq:projCreneauApprox} to the roughness positioned at $s=0.008$ has a maximum mean fluctuation rate of $2.2\times10^4$ against $1.6\times10^3$ and $3.5\times10^2$ for the roughness at $s_0=0.016$ and $s_0=0.056$ respectively.
The perturbation triggered by the roughness located at $s_0=0.008$ calculated by solving the LNSE is represented in figure \ref{fig:creneau3D_bouge}. The only notable difference from the perturbation calculated with the approximation \eqref{eq:projCreneauApprox} is in the region of the roughness highlighted by the yellow ellipse, where the response exhibits a spatial structure similar to the one of the roughness. Note that a PSE method would not be able to compute the perturbation in the vicinity of the roughness since it does not capture the non-modal mechanisms triggered by the roughness (which are well reproduced when solving the LNSE (resolvent) and accounted in the resolvent analysis by the sub-optimal modes).

\subsubsection{Case of a compact roughness in both the curvilinear $s$- and spanwise $z$-directions}\label{results:ellipse}

Considering again a roughness in separate form given by equation \eqref{eq:separate}, we now choose:
\begin{eqnarray}
    {h}_s(s)&=&\sqrt{1-s^2} \textrm{ for } s\in[-1,1] \;\;\;\;\textrm{ and }\;\;\;\; {h}_s(s)=0 \textrm{ for } |s|>1 \\
    {h}_z(z)&=&e^{-z^2}.
\end{eqnarray}

\begin{figure}[h!]
\begin{center}
\begin{tabular}{rlrl}
(a) & & (b) & \\
& \hspace{-0.5cm}\includegraphics[trim =02mm 20mm 11mm 40mm, clip,width=0.45\textwidth]{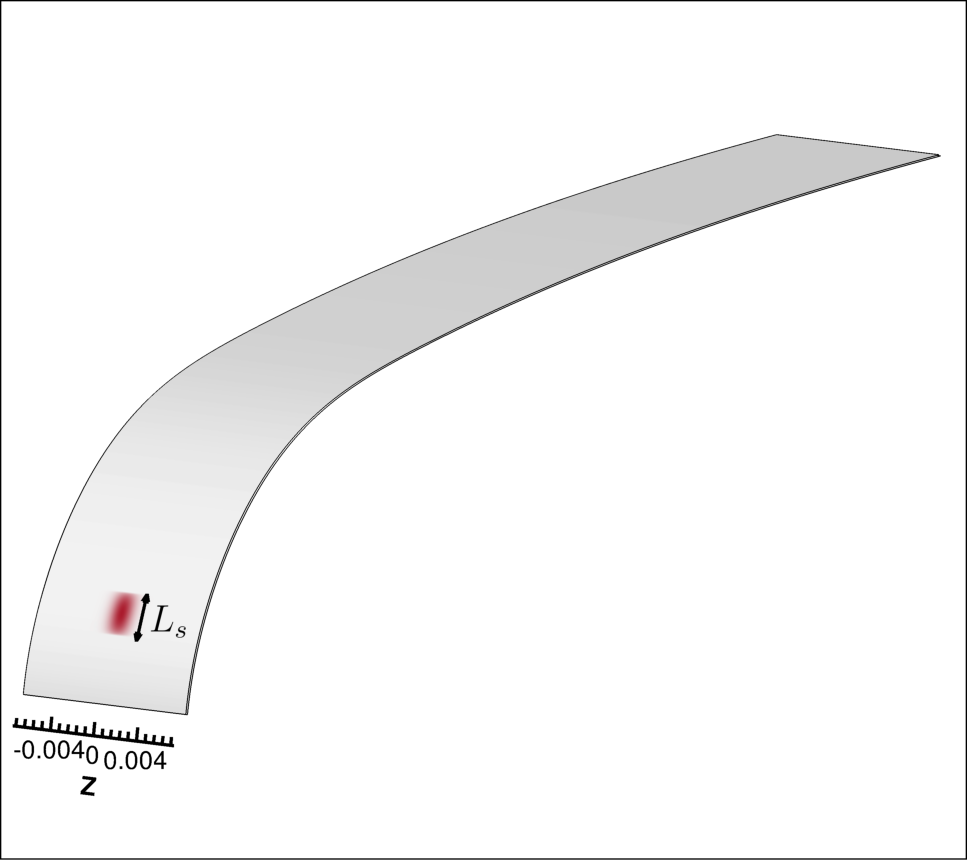} & & \hspace{-0.5cm}\includegraphics[trim =02mm 12mm 2mm 62mm, clip,width=0.45\textwidth]{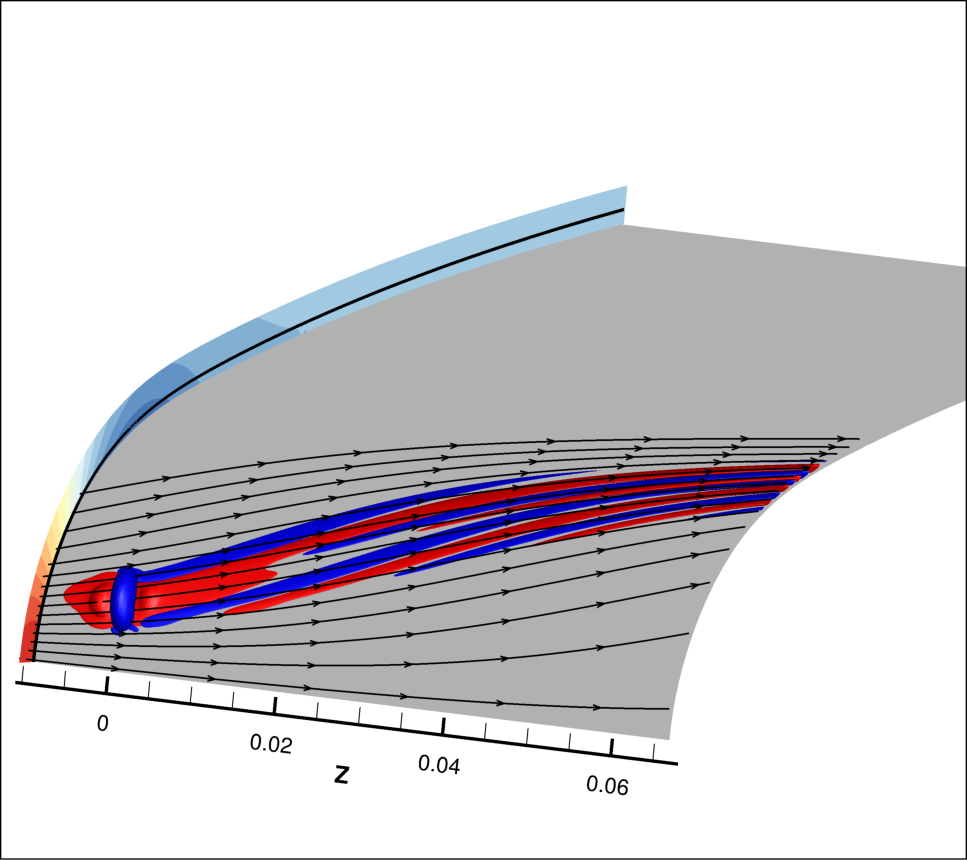}
\end{tabular}
\end{center}
 \vspace{-0.2cm}
\caption{$(a)$: 3D representation of the localised roughness $h(s,z)/H$ with $s_0=0.008\approx 82 \Delta$, $L_s/\Delta=20.6$, $L_z/\Delta=10.3$. $(b)$: Spatial structure of the corresponding response calculated by solving the LNSE, with the same representation as in figure \ref{fig:Response3D}.}
\label{fig:ellipse3D}
\end{figure}

 Hence, the roughness is a Gaussian in the $ z $-direction localised around $ z=0 $ and a semi-ellipse localised at $s_0$ in the $ s$-direction with the major axis being $ L_s $ and the minor axis $ He^{-z^2}/2$.
We set $s_0=0.008 \approx 82 \Delta$, $L_s/\Delta=20.6$, $L_z/\Delta=10.3$. % and $H/\Delta=0.02$.
The roughness shape $ h/H $ is represented in figure \ref{fig:ellipse3D}$(a)$ and can be considered as representative, for example, of the presence of an insect on a wing surface. The spatial structure of the response calculated solving the LNSE is shown in figure \ref{fig:ellipse3D}$(b)$. The response develops from the position of the roughness and is convected downstream in the direction of the external streamlines where it acquires the spatial structure of a cross-flow mode.

 With equation \eqref{eqn:Fourier}, the wall displacement reads
$ {h}(s,z)/H=(1/2){h}_s((s-s_0)/L_s)\int_{-\infty}^{+\infty}
     \hat{h}_{z,\beta}e^{i\beta z} \; d\beta $
where $ \hat{h}_{z,\beta}$ is the Fourier transform of a Gaussian wave, so that $ \hat{h}_{z,\beta}=(L_z/2\sqrt{\pi}) \exp(-(\beta L_z)^2/4) $. 
%We can analytically calculate the Fourier transform $c_z$ of $\hat{h}_z$:
% \begin{equation}
%     c_z(\beta)=\frac{1}{2\pi}\int_{-\infty}^{\infty}e^{-\alpha z^2}e^{-i\beta z} dz=\frac{1}{2\sqrt{\pi\alpha}}e^{\frac{-\beta^2}{4\alpha}}
% \end{equation}
  We then calculate the response of the system by using equation \eqref{eq:ProjBasis}. We will, once again, make the assumption that the first singular value is sufficiently dominant to neglect the following ones. Moreover, only the roughness with spanwise wavenumber $|\beta\Delta|<0.5$ are taken into account:
\begin{equation}\label{eq:ResponseEllipseApprox}
\frac{\boldsymbol{u}(x,y,z)}{H}\approx \int_{0}^{\frac{0.5}{\Delta}}\Re\left(\hat{h}_{z,\beta}\sigma_{1,\beta}\left\langle\hat{h}_{1,\beta},{h}_s\left(\frac{s-s_0}{L_s}\right)\right\rangle_w\boldsymbol{\hat{u}}_{1,\beta}(x,y)e^{i\beta z}\right)\;d\beta,
  \end{equation}
 and the integral is discretised with the extended Simpson's rule with steps of $ \Delta\beta=0.001$.
 
\begin{figure}
\centering
\includegraphics[trim =00mm 0mm 0mm 2mm, clip,width=0.7\textwidth]{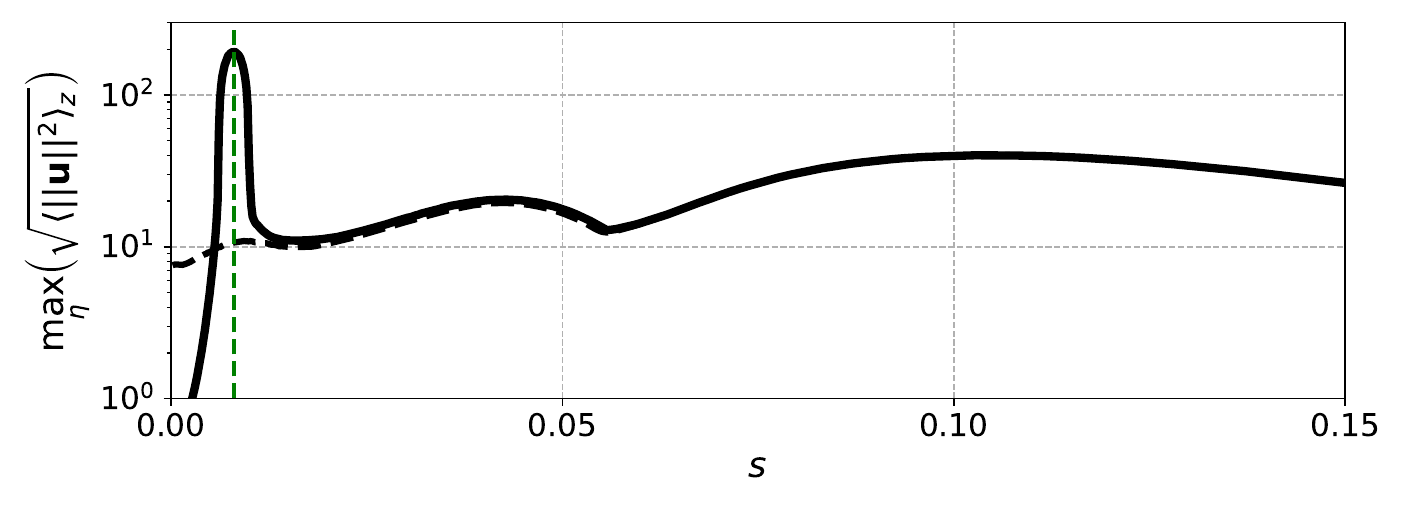}
\caption{Curvilinear evolution of $\max_{\eta}\left(\sqrt{\langle\|\boldsymbol{ u}\|^2\rangle_z} \right)$ of the response computed using equation \eqref{eq:ResponseEllipseApprox} (dashed line) and the exact equation \eqref{ResolventDiscret} (solid line). The location of the centre of the roughness is shown by the green vertical dashed line.}
\label{fig:reponseEllipse}
\end{figure}

 In figure \ref{fig:reponseEllipse} is represented the curvilinear evolution of the wall normal maximum of the mean fluctuation rate $\max_{\eta}(\sqrt{\langle\|\boldsymbol{ u}\|^2\rangle_z})/H$. The response calculated from equation \eqref{eq:ResponseEllipseApprox} is plotted with a dashed line while the response computed from the resolution of \eqref{ResolventDiscret} is drawn with a solid line. We observe, once again, a good agreement between both responses downstream of the roughness, with a second amplification at $s=0.054$ and two local maxima at $s=0.042$ and $s=0.1$ reaching respectively $19.5$ and $41$. As in figure \ref{fig:CreneauPartiel}, the fact that the maximum response magnitude is reached at the location of the roughness is related to the low energy of the response due to the strong stability of the flow configuration considered in this paper.

 %We notice that, due to the superposition of the optimal responses of different $\beta$ values, the spatial structure of the perturbation exhibits sharp variations at the attachment-line with an orientation reminiscent of an attachment-line mode, in the area where its magnitude is low. However, it recovers the features of a cross-flow mode downstream where the magnitude of the response is the highest. A CHANGER

\subsubsection{Case of chordwise and spanwise harmonic roughness}\label{results:harmonics}

In this section, we consider a roughness of harmonic shape:
\begin{equation}
\frac{h(s,z)}{H}= \underbrace{\cos(\gamma s)}_{{h}_\gamma(s)}\cos(\beta z),    
\end{equation}
 with $(\gamma,\beta) \in\mathbb{R}^2$.
  The exact and approximated responses are respectively given by:
 \begin{eqnarray}
\frac{\boldsymbol{u}(x,y,z)}{H}&=&
 \Re( \boldsymbol{\hat{u}}(x,y)e^{i\beta z}) \;\;\;\;\;\; \mbox{with} \;\;\;\;\;\; \boldsymbol{\hat{u}}=R_\beta \boldsymbol{{h}}_\gamma \label{eq:cosDirect}\\
 &\approx&
\Re\left(\sigma_{\beta,1}\langle\hat{h}_{\beta,1}(s),\cos(\gamma s)\rangle_w\boldsymbol{\hat{u}}_{\beta,1}(x,y)e^{i\beta z}\right) \label{eq:projCosApprox}
%\\&=&
%\frac{1}{4}\sigma_{\beta,1}\boldsymbol{\hat{u}}_{\beta,1}(x,y)\int_{-\infty}^\infty \left(\overline{\hat{h}_{\beta,1}(s)}\cos(\gamma s) ds\right) e^{i\beta z}+\mbox{c.c.}
%\\&=&
%\frac{1}{4}\sigma_{\beta,1}\boldsymbol{\hat{u}}_{\beta,1}(x,y)\int_{-\infty}^\infty \left(\overline{\hat{h}_{\beta,1}(s)}e^{i\gamma s} ds\right) e^{i\beta z}+\mbox{c.c.}
%\\&=&
%\frac{2\pi}{4}\sigma_{\beta,1}\boldsymbol{\hat{u}}_{\beta,1}(x,y)\overline{\frac{1}{2\pi}\int_{-\infty}^\infty \left({\hat{h}_{\beta,1}(s)}e^{-i\gamma s} ds\right)} e^{i\beta z}+\mbox{c.c.}
=
\Re\left(\pi \sigma_{\beta,1} \overline{\hat{\hat{h}}_{\beta,1}(\gamma)}\boldsymbol{\hat{u}}_{\beta,1}(x,y) e^{i\beta z}\right),
\end{eqnarray}
  where $\hat{\hat{h}}_{\beta,j}(\gamma)=\frac{1}{\pi}\int_{0}^\infty {\hat{{h}}_{\beta,j}(s)}e^{-i\gamma s} ds$ is the Fourier transform of the $ j^{th}$ symmetric optimal roughness  $\hat{{h}}_{\beta,j}(s)$.

\begin{figure}
\centering
\includegraphics[trim =00mm 0mm 0mm 0mm, clip,width=0.7\textwidth]{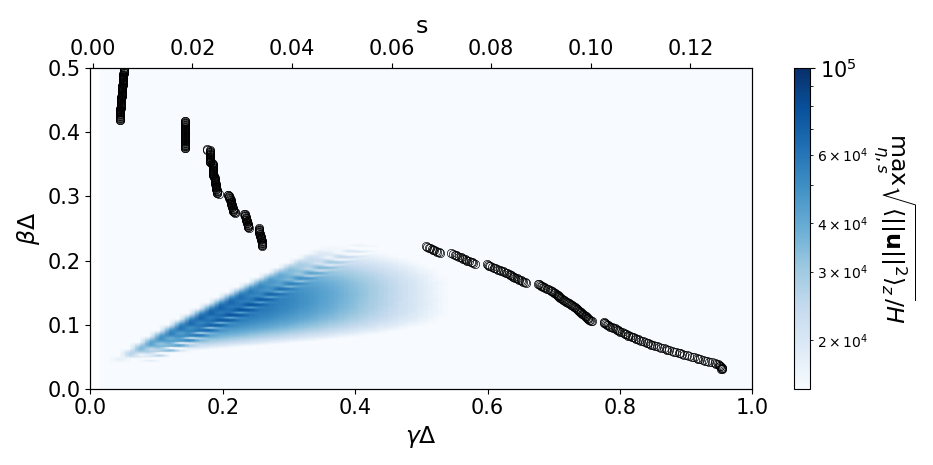}
\caption{Representation, according to $\beta\Delta$ and $\gamma\Delta$, of $\max_{\eta,s}\left(\sqrt{\langle\|\boldsymbol{ u}\|^2\rangle_z} \right)/H$. The curvilinear location where the maximum of the function for optimal perturbations is reached is plotted in black circles with respect to the top axis.}
\label{fig:map}
\end{figure}

  In figure \ref{fig:map} is represented, for each couple $(\beta\Delta,\gamma\Delta)\in([0,0.5]\times[0,1])$, the value of $\max_{\eta,s}\left(\sqrt{\langle\|\boldsymbol{ u}\|^2\rangle_z} \right)/H$ using equation \eqref{eq:cosDirect}.
The same representation as figure \ref{fig:map} but considering equation \eqref{eq:projCosApprox} was also computed (not shown here) and gave similar results. This validates the approximation made in equation \eqref{eq:projCosApprox} and reveals that when the roughness is not localised in the curvilinear direction, the maximum amplitude of the response is well calculated by the projection on the optimal roughness, and this for all values of $\beta$ and $\gamma$. Moreover, in the case of perturbations computed with the approximation \eqref{eq:projCosApprox}, it is straightforward that the curvilinear position of the maximum represented in figure \ref{fig:map} is the same for all values of $\gamma$ and corresponds to the maximum of the optimal perturbation at the value of $\beta$ considered. Thus, in figure \ref{fig:map} are represented in black circles the curvilinear position at which the optimal perturbation reach $\max_{\eta,s}\left(\sqrt{\langle\|\boldsymbol{\hat{u}}_{\beta ,1}\|^2\rangle_z} \right)/H$ for each value of $\beta\Delta>0.03$.
  
%\begin{figure}
%\centering
%\includegraphics[trim =04mm 65mm 32mm 40mm, clip,width=0.55\textwidth]{Images/creneau3D_bouge_kz0.11_Rer25000_Res300_pulsSpat0.34.png}
%\caption{3D representation of the studied chordwise and spanwise harmonic wave roughness shape $ h(s,z)/H$. The negative and positive values are depicted in blue and red respectively. ADAPT FIGURE. ADD THE RESPONSE (EXACT COMPUTATION)?}
%\label{fig:wave3D}
%\end{figure}

The spanwise wavenumbers $\beta\Delta$ associated with high values are between $0.05$ and $0.25$, which is directly related to the singular values $\sigma_1$ represented in figure \ref{fig:SV}. For a given spanwise wavenumber, the magnitude of the response is linked to the value of $\lvert \hat{\hat{h}}_{\beta,1}(\gamma)\rvert$. For example, for $\beta\Delta=0.11$, a first maximum is reached at $\gamma\Delta=0.18$ even if it does not correspond to the curvilinear wavenumber of the optimal roughness at its maximum magnitude ($k_s\Delta=0.32$ at $s=0.008$). However, this high value is justified by the fact that, as represented in figure \ref{fig:ksDelta_ForcingOpt_kz0.11}, $\gamma\Delta=0.18$ is close to the curvilinear wavenumber of the optimal roughness over a large range of $s$ values, including at positions where the roughness magnitude is high. Although $k_s\Delta=0.32$ corresponds to the curvilinear wavenumber of the optimal roughness at its maximum magnitude, the strong variation of $k_s\Delta$ at $s=0.008$ justifies the lower value of $\lvert \hat{\hat{h}}_{0.11/\Delta,1}(0.32/\Delta)\rvert$. 

As the value of $\beta$ increases, the most critical $\gamma$ values also increase. This is explained by the fact that, as shown in figure \ref{fig:s_Psi_kz}$(a)$, the position of the magnitude maximum shifts downstream with increasing $\beta$, while remaining lower than $s=0.016$ reached for $\beta\Delta=0.3$. For increasing $ s $, the direction of the external streamlines gets closer to the chordwise direction. Since the $\Psi$ angle at the position of the maximum magnitude of the optimal roughness remains constant with increasing $\beta$ (figure \ref{fig:s_Psi_kz}$(b)$), the roughness also has a direction that approaches the chordwise direction at the location of its maximum magnitude. This implies an increase of $k_s$ at this position at least proportional to that of $\beta$, which explains the increase of the most critical $ \gamma $ values in figure \ref{fig:map}. Moreover, the position of the maximum varies from $s\approx0.005$ for $\beta\Delta=0.5$ to $s=0.126$ for $\beta\Delta=0.03$. This evolution is slightly different from the one represented in figure \ref{fig:s_Psi_kz} because we evaluate the maximum of different quantities. The ``jumps'' that can be observed between different values of $\beta\Delta$ despite the fact that the spatial structure of the optimal perturbations evolves continuously with $\beta$ is related to the presence of ``plateau" in the curve of $\max_{\eta}\left(\sqrt{\langle\|\boldsymbol{\hat{u}}_{\beta ,1}\|^2\rangle_z} \right)(s)$ . The jump at $\beta\Delta=0.22$ is linked to the switch from a position corresponding to the first amplification to a location related to the second amplification.

%\begin{comment}
%The knowledge of $\hat{\hat{h}}_{\beta,1}(\gamma)$ is all the more important since any real even roughness ${h}_s(s)$ can be rewritten as ${h}_s(s)=\int_{-\infty}^{\infty}\hat{h}_s(\gamma)e^{i\gamma s}d\gamma=2\int_{0}^{\infty}\hat{h}_s(\gamma)\cos(\gamma s)d\gamma$ %\underbrace{=}_{\hat{h}_s(s)\in\mathbb{R}}\frac{1}{\pi}\int_{0}^{\infty}c_s(\gamma)cos(\gamma s)d\gamma$
%and projecting this roughness onto an optimal roughness $\hat{h}_1$ writes:
%\begin{equation}
%\langle\hat{h}_{\beta,1}(s),{h}_s(s)\rangle_w=2\int_{0}^{\infty}\overline{\hat{\hat{h}}_{\beta,1}}(\gamma) \hat{h}_s(\gamma)\;d\gamma
%\end{equation}
%and the possibility of considering only a singular value was validated in the previous sections.
%Therefore, as was the case with the singular values for the $z$-direction, knowing the values of $\hat{\hat{h}}_{\beta,1}(\gamma)$ gives us intrinsic information about the most amplified curvilinear wavenumbers.
% \end{comment}
 
 The overall picture indicates that the most dangerous roughness are characterised by $ \beta \leq 0.5 \gamma$, hence rectangular trellis elongated in the $ z$-direction by an aspect ratio of at least $2$.
 The characteristic size of these roughness is around $ \beta \Delta \approx 0.1$, that is $ L_z/\Delta \approx 60 $.

These calculations can be used to make a first attempt to predict the roughness height $H_c$ for which nonlinear effects may appear. Some studies suggest that the first nonlinear effects appear in a 3D flow when the magnitude of the disturbance velocity reaches about $10\%$ of the baseflow velocity \citep{arnal2008practical,tempelmann2012swept2}. 
Thus, considering the criterion: $\max_{\eta,s}\sqrt{\langle\|\boldsymbol{u}\|^2\rangle_z}/\sqrt{2}=0.1U^\infty$, we obtain, for a roughness with $\gamma \Delta=0.18$, a critical height value of $H_c=4.92\times10^{-6}\approx4.9\%\Delta$ for $ \beta \Delta=0.11$.  It corresponds to the order of magnitude of the wall defects that can be encountered on real wings \citep{Radeztsky1999}. This is therefore a first indication that we could see non-linear effects appear before the end of the domain in these flow conditions for unpolished aerofoils. For a roughness of height $H=4.92\times10^{-6}$, the criterion is reached for $s=0.1$. It is important to note that this roughness size is small enough to allow the linearisation of the wall boundary condition to remain valid. Indeed, in our case, the displacement thickness $\delta^*$ is greater than $10^{-4}$ and the approximation remains valid for roughness of height below $5\%\delta^*$ \citep{schrader2009receptivity,tempelmann2012swept2}.

%In figure \ref{fig:hcrit_ks} are represented the critical heights $H_c$ associated with harmonic wave roughness of varying $\gamma$ and calculated with the projection method. We observe that the lowest $H_c$ values are not reached for $k_s\Delta=0.32$ but rather for $k_s\Delta\approx0.2$ values.

%\begin{figure}
%\hspace{1.5cm}\includegraphics[trim =-0mm 0mm 0mm 0mm,width=0.8\textwidth]{hcrit_ks.pdf} 
%\caption{Evolution of the critical height of a non-localised square wave roughness according to its curvilinear wavenumber. I WOULD RATHER PLOT A GRAPH IN THE $ (\gamma \Delta, \beta \Delta) $ PLANE OF ISOVALUES OF $ H_c $.  \textcolor{blue}{Changer $\tilde{f}$ en $H$}}
%\label{fig:hcrit_ks}
%\end{figure}

\section{Conclusion}

In this paper, we studied the receptivity of a swept aerofoil to wall roughness using wall displacement based resolvent analysis. An incompressible flow covering the whole leading edge was considered using a global framework. 
We have identified the most critical spanwise and curvilinear wavenumbers, as well as the spatial structure of the optimal roughness and associated perturbations. In the case of the spanwise wavenumber with the highest dominant singular value, the optimal response corresponds to a cross-flow mode with two local maxima. The optimal roughness is located near the attachment-line with a wavevector nearly orthogonal to the external streamlines.
Moreover, the increase of the spanwise wavenumber shifts the location of the optimal roughness further downstream and the location of the optimal response closer to the attachment-line. The most amplified curvilinear wavenumbers rise with increasing spanwise wavenumber. For all spanwise wavenumbers, the optimal roughness has a maximum magnitude close to the attachment-line, confirming the results of previous studies \citep{tempelmann2012swept1,Radeztsky1999,thomas2017predicting}.
When the resolvent operator is low-rank (few dominant singular values dominate), the optimal roughness and associated responses provide a low-order model. Once the optimal roughness and perturbations have been computed, the low-order model can be used to compute the flow response to any small amplitude roughness shape at a reduced computational cost compared to the direct resolution of the LNSE. 

Responses to various roughness (periodic and compact in both chordwise and spanwise directions) were finally computed. We verified that the magnitude of the response is the highest when the roughness has a Fourier transform involving spanwise wavenumbers corresponding to the highest singular values and curvilinear wavenumbers representative of the optimal roughness over a large range of $s$ where its amplitude is high. 
Approximate responses calculated with a single singular value were compared to the responses obtained from the resolution of the LNSE and showed good agreement downstream of the roughness. When the roughness is localised in the curvilinear direction, a peak of magnitude can appear at the location of the roughness that cannot be captured by the low-rank method. %, but this peak is only present for low energy perturbations. 
In particular, this shows that the knowledge of the spatial structure of some optimal perturbations allows us to predict the characteristics of the response to a wide range of roughness.
A first attempt to predict the height of a non-localised square wave roughness required to trigger nonlinear effects before $X=15\%$ was made. A critical roughness height around $10^{-6}$ was found for a square wave roughness with a spanwise wavenumber matching the one of the most amplified optimal roughness. This size being of the order of the wall defect sizes that can be found on realistic aerofoils, it is a first indication of the possible transition before $X=15\%$ in this particular flow conditions. 
One possible prospect of this work is to compare the results of the resolvent analysis with experimental or DNS results.

%Perspectives for control.
%A control strategy studied by \citet{saric2006stability} is, with the help of an alley of micro-roughness, to generate “killer” cross-flow waves which will, by non-linear interaction, modify the baseflow and thus potentially delay the transition by making the target wave less amplified. In their case, the wavelength ratio leading to the most effective control is $\lambda_{z,k}/\lambda_{z,t}=8/12=2/3$. This ratio depends on the profile and the Reynolds number considered but is generally $1/2$ or $2/3$.  An extension would be to consider the fully nonlinear framework of \citet{rigas2021nonlinear} to rigorously optimise the roughness that generate the killer-waves.

\clearpage

\begin{Backmatter}
\section*{Acknowledgment}
The authors thank Jean-Philippe Brazier for providing the local stability analysis code.

\paragraph{Funding Statement}
The study was supported by a grant from the Agence de l’innovation de défense (Defence Innovation Agency).

\paragraph{Declaration of Interests}
The authors declare no conflict of interest.

\paragraph{Data Availability Statement}
Further details on underlying data are available from the corresponding author (E.K.).

\paragraph{Ethical Standards}
The research meets all ethical guidelines, including adherence to the legal requirements of the study country.

\bibliographystyle{jfm}

\begin{bibliography}{references}
\end{bibliography}
\end{Backmatter}

\end{document}